\RequirePackage{fix-cm}
\documentclass[smallextended]{svjour3}       
\smartqed  
\usepackage{bchart}
\usepackage{caption}
\usepackage{amsmath,amssymb,amsfonts}
\usepackage{graphicx}
\usepackage{textcomp}
\usepackage{color,soul}
\usepackage{multirow}
\usepackage{threeparttable}
\usepackage[]{mdframed}
\usepackage[toc,page]{appendix}
\usepackage{csvsimple}
\usepackage{longtable}

\newcommand{\secref}[1]{Section~\ref{sec:#1}}
\newcommand{\tabref}[1]{Table~\ref{tab:#1}}
\newcommand{\secsref}[2]{Sections~\ref{sec:#1}-\ref{sec:#2}}

\newcommand{\figref}[1]{Figure~\ref{fig:#1}}
\newcommand{\eqaref}[1]{Equation~\eqref{eq:#1}}

\newcommand{\appenref}[1]{Appendix~\ref{appendix:#1}}

\begin{document}


\title{Predicting the Understandability of Computational Notebooks through Code Metrics Analysis}

\titlerunning{Predicting Computational Notebook Understandability through ... }        

\author{Mojtaba Mostafavi Ghahfarokhi\and
        Alireza Asadi \and
        Arash Asgari \and
        Bardia Mohammadi \and
        Abbas Heydarnoori \and
        Masih Beigi Rizi
}

\institute{M. Mostafavi Ghahfarokhi\at
              Department of Computer Engineering, Sharif University of Technology \\
              \email{m.mostafavi@sharif.edu}           
            \and
            A. Asadi \at
            Department of Computer Engineering,
            Sharif University of Technology \\
                \email {alireza.asadi@sharif.edu}
            \and
            A. Asgari \at
            Department of Computer Engineering, Sharif University of Technology \\
                \email{arash.asgari@sharif.edu}
            \and
            B. Mohammadi \at
            Department of Computer Engineering, Sharif University of Technology \\
                \email{bardia.mohammadi@sharif.edu}
            \and
            A. Heydarnoori \at
            Department of Computer Engineering, Sharif University of Technology \\
                \email{heydarnoori@sharif.edu \newline A. Heydarnoori is currently affiliated with the Department of Computer Science at Bowling Green State University, USA.}
            \and
            M. Beigi Rizi \at
            Department of Computer Engineering, Sharif University of Technology \\
                \email{masih.beigi@sharif.edu}
}

\date{Received: date / Accepted: date}


\maketitle

\begin{abstract}
Computational notebooks have become the primary coding environment for data scientists. Despite their popularity, research on the code quality of these notebooks is still in its infancy, and the code shared in these notebooks is often of poor quality. Considering the importance of maintenance and reusability, it is crucial to pay attention to the understandability of the notebook code and identify the notebook metrics that play a significant role in
its understandability. The level of code understandability is a qualitative variable closely associated with the user's opinion about the code. Traditional approaches to measuring it either use limited questionnaires to review a few code pieces or rely on metadata such as likes and votes in software repositories. In our approach, we enhanced the measurement of the understandability level of Jupyter notebooks by leveraging user opinions related to code understandability within a software repository. As a case study, we started with 542,051 Kaggle Jupyter notebooks, compiled in a dataset named DistilKaggle, which we introduced in our previous research. To identify user comments associated with code understandability, we utilized a fine-tuned DistilBERT transformer. We established a \emph{user-opinion-based criterion} for measuring code understandability by considering the number of code understandability-related comments, the upvotes on those comments and the total views of the notebook received by the notebook. We refer to this criterion as User Opinion Code Understandability (UOCU), which has been proven to be much more effective than previous approaches. A hybrid approach combining UOCU with total upvotes further improved this criterion.
Additionally, we trained machine learning models to classify notebook understandability solely based on notebook metrics. We collected 34 metrics for 132,723 final notebooks using the hybrid approach criterion. Our predictive model, built using a Random Forest classifier, achieved 89\% accuracy in classifying code understandability levels in computational notebooks.

\end{abstract}

\section{Introduction}
\label{sec:Intro}
Computational notebooks have become the primary coding environment for data scientists~\cite{psallidas2019data}. They offer several advantages over traditional software development environments, including easy documentation, code sharing, result analysis, visual and intuitive code development, and the ability to compile, execute, and modify code cells without re-executing the entire notebook. The advantages outlined above establish computational notebooks as a fundamental tool in the development of data analysis solutions. For instance, a case study at Microsoft highlighted that data scientists rely on computational notebooks for sharing and reusing their code across software teams~\cite{epperson2022strategies}. In another empirical study, cleaned computational notebooks were introduced as a best practice for a frictionless transition of data-driven projects from the exploration stage to production as a software engineering product~\cite{quaranta2022assessing}.

The quality and understandability of implementations in computational notebooks are crucial for various purposes, such as education to provide coding best practices for future data scientists~\cite{wang2020better}, notebook reusability~\cite{epperson2022strategies}, and maintainability~\cite{wang2020assessing,yang2021subtle,dong2021splitting}. Recent studies have focused on improving the quality of computational notebooks through enhanced documentation~\cite{liu2019understanding,rule2018exploration,wang2022documentation,yang2021subtle,venkatesh2023enhancing}, notebook structure~\cite{wenskovitch2019albireo,titov2022resplit,jiang2022elevating}, and managing notebook variants and revisions~\cite{subramanian2019supporting,jiang2022elevating}. However, efforts to enhance code understandability in computational notebooks are still in the early stages~\cite{venkatesh2021automated}. Our study aims to address this problem.

Code Understandability (CU), also referred to as program understandability or code comprehensibility, is the active process by which developers acquire knowledge about a software system through exploring software artifacts, reading source code, and consulting documentation~\cite{xia2017measuring}. On average, developers spend approximately 60\% of their time engaged in program comprehension during software development and maintenance~\cite{xia2017measuring}. Various approaches have been proposed to measure CU, often involving gathering feedback from individuals regarding a program or specific code segments. These studies employ different response variables, including time, accuracy, opinion, and visual metrics, as criteria for assessing CU. Some studies directly solicit opinions from a limited number of developers about specific code pieces~\cite{sykes1983effect,buse2009learning,medeiros2018investigating,scalabrino2019automatically}. Other studies rely on metrics such as likes, stars, or votes on software repositories to gauge CU~\cite{lu2018internal,nasehi2012makes,wang2022documentation,liu2021haconvgnn}. However, findings from small-scale studies may not be widely applicable or provide a comprehensive understanding of how developers perceive code. Additionally, user votes on software projects may be influenced by factors beyond CU, such as ease of use, algorithm accuracy, popularity, and marketing considerations.

In our study, we have employed the \emph{opinion response variable}~\cite{oliveira2020evaluating} to enhance the measurement of CU in a unique manner. Instead of conducting surveys, direct inquiries, or relying solely on user votes, we have developed a novel approach that involves extracting and analyzing comments regarding the understandability of notebooks in a repository. As a case study, we chose Kaggle\footnote{https://www.kaggle.com}, which is a notebook repository with a variety of interactive and collaborative functionalities. In this regard, we began by obtaining the metadata files of the Jupyter notebooks and then applied preprocessing to the user comments. We utilized DistilBERT\footnote{DistilBERT is designed to be smaller, faster, and more efficient while maintaining a high level of performance in natural language processing tasks. It is particularly effective for understanding the contextual semantics of user comments, which is crucial for evaluating CU in computational notebooks.}~\cite{sanh2019distilbert}, a distilled version of the BERT (Bidirectional Encoder Representations from Transformers) model. Using a fine-tuned DistilBERT transformer, we identified user comments related to CU and provided a metric considering how much each comment attracted other users' attention. We named this metric User Opinion\footnote{What we mean by ``opinion'' refers to user comments, and it has no relation to high upvotes or other metadata.} Code Understandability (UOCU). Compared to a dataset labeled by 42 experts, UOCU has been shown to outperform current state-of-the-art methods for measuring the understandability of notebooks, as demonstrated in our evaluation. Our findings indicate that a hybrid metric combining UOCU and notebook upvotes is the most effective for assigning CU scores and supervising CU in machine learning models.


Another key contribution of our study is the identification of notebook-specific metrics that exert a more substantial influence on CU. While prior research has established various code metrics (e.g., lines of code, maximum line length, and Cyclomatic Complexity) for evaluating CU in conventional programming languages such as Java, Python, and C~\cite{buse2009learning,posnett2011simpler,kasto2013measuring,scalabrino2019automatically,grotov2022large}, many of these metrics (e.g., coupling, cohesion, and depth of the inheritance tree) are not directly applicable to notebooks. Notably, less than 3.5\% of notebooks incorporate classes or object-oriented programming principles, rendering the depth of inheritance tree and other class-based CU metrics ineffective for CU evaluation in Jupyter notebooks. Due to the structural differences between notebooks and traditional code scripts, additional metrics may be relevant, as highlighted in recent studies~\cite{grotov2022large,wang2020assessing,wang2020better,titov2022resplit,ono2021interactive,wang2021restoring,zhu2021restoring,dilhara2021understanding,titov2022resplit,yang2021subtle,wang2022documentation,venkatesh2023enhancing}. Moreover, while some studies propose certain notebook metrics as promising for assessing CU, to the best of our knowledge, based on our literature review, they lack large-scale evaluation~\cite{wang2022documentation}, unlike our study, which evaluates CU in more than 500k notebooks. To address this gap, we identified metrics relevant to CU, building on our previous work~\cite{mostafavi2024DistilKaggle}, which introduced DistilKaggle—a dataset distilled from 542,051 notebooks shared on the Kaggle platform—by calculating 34 code metrics for each notebook.


Using this dataset of notebook metrics and the best CU score we identified, we trained a machine learning model using four algorithms: CatBoost, XGBoost, Decision Tree, and Random Forest.
The Random Forest algorithm yielded the highest accuracy for notebook CU classification, achieving an accuracy of 89\% and an F1-score of 88\%, improving previous criteria. With this acceptable level of accuracy and F1-score, we identified the most influential notebook metrics for CU. These metrics included \emph{developer expertise level, number of blank lines of code, number of visualization data types, number of executed cells, number of H1 tags in markdown headlines} and \emph{number of lines of markdown cells}. Additionally, by applying machine learning algorithms to different groups of code metrics, we concluded that notebook-specific metrics have the greatest impact on code understandability compared to more general metrics.

\noindent In summary, our contributions are as follows:
\begin{itemize}

 \item introducing a new criterion for CU of computational notebooks

 \item presenting a ground truth and creating a model to classify CU levels within Jupyter notebooks

 \item identifying the most impactful notebook and code metrics for improving the CU

\end{itemize}

This paper is organized as follows: \secref{motivating} presents a motivating example, and \secref{metrics} provides a review of notebook metrics. \secref{problem} details the problems and research questions addressed in this study. Our proposed approach is outlined in \secref{approach}, and \secref{evaluation} presents our evaluations and results. \secref{discussion} discusses the results and their potential applications, while \secref{related_work} covers related work. The limitations and validity risks of our findings are explained in \secref{threads}. Finally, \secref{conclusion} concludes the paper and offers directions for future research.

Our  dataset  and  experimental  source  code are available online\footnote{https://github.com/ISE-Research/NotebookCU}.



\section{Motivating Example}
\label{sec:motivating}
Some studies leverage the quantity of user votes received by a code or notebook within a software repository as an indicator of its understandability level~\cite{nasehi2012makes,lu2018internal,wang2022documentation,liu2021haconvgnn}.
However, user votes can be influenced by various factors unrelated to CU, such as output accuracy, notebook popularity, or ease of reuse.

For example, notebooks producing visually appealing results or addressing popular topics might garner more upvotes even if their code is not well-organized or documented. To explore this further, we selected 10 notebook samples from Kaggle, along with their upvote data up to August 10, 2023(latest upvote in \tabref{notebookvote}).

Three authors of this manuscript, with expertise in Jupyter notebooks and familiarity with code comprehension research, evaluated each sample, categorizing it as either normal or good CU\footnote{Following this assessment, they developed an annotation guideline to clearly define the concepts and criteria for CU classification, which supports this research and is provided in \appenref{guidline2}.}. The average results of their evaluations, presented in \tabref{notebookvote}, represent their collective assessments across the notebook samples.
The first four notebooks in this table, despite having higher upvotes and an average of 23,000 views, have fewer comments related to CU. In contrast, there are notebooks that, although they have high upvotes and an average of 2,500 views, contain significant comments indicating a good understanding of the notebooks by the readers. These notebooks were also labeled as having good CU by our annotators. Here are some of the comments: ``Nicely organized file and good EDA'', ``Well-explained notebook Very well presented'', ``nice kernel, I enjoyed reading your work'', ``great job Outstanding'', ``neatly, well-explained Kaggle Notebook'', ``Awesome notebook with impressive visualizations'' and ''Excellent work with many useful charts''. This observation suggests that user opinions may serve as better indicators of code understandability than upvotes alone.

\begin{table}[h]
  \caption{Examples of Notebooks Upvotes and Experts' Scores}
  \centering
  \label{tab:notebookvote}
  \begin{tabular}{c c c c}
    \hline
    \textbf{Notebooks*}  & \textbf{Latest Upvote}& \textbf{Latest View} & \textbf{Expert Score} \\
    \hline
    \texttt{\textbf{A}} &66 & 40125&  normal\\
    \hline
    \texttt{\textbf{B}} &72 & 12009& normal\\
    \hline
    \texttt{\textbf{C}} &96 & 7551& normal\\
    \hline
    \texttt{\textbf{D}} &113 & 10650 & normal\\
    \hline
    \texttt{\textbf{E}}  &5 &1010 &good\\
    \hline
    \texttt{\textbf{F}}  &21 &4550 &good\\
    \hline
    \texttt{\textbf{G}}  &21 &1514& good\\
    \hline
    \texttt{\textbf{H}}  &40 &3251 &good\\
    \hline
  \end{tabular}

\footnotesize{* Links of notebooks:}
\begin{itemize}
  \item[-] \textbf{A}:https://www.kaggle.com/code/panamby/loan-default-prediction
  \item[-] \textbf{B}:https://www.kaggle.com/code/christofhenkel/weighted-kappa-loss-for-keras-tensorflow
  \item[-] \textbf{C}:https://www.kaggle.com/code/bicotsp/pmtest1
  \item[-] \textbf{D}:https://www.kaggle.com/code/raddar/paris-madness
  \item[-] \textbf{E}:https://www.kaggle.com/code/naren204/speed-at-which-pandemic-covid-19-spreading
  \item[-] \textbf{F}:https://www.kaggle.com/code/avnika22/lgbm-predicting-house-prices
  \item[-] \textbf{G}:https://www.kaggle.com/code/nikhileshkos/play-store-analysis
  \item[-] \textbf{H}:https://www.kaggle.com/code/landfallmotto/60k-stack-overflow-questions-keras-lstm-and-cnn

\end{itemize}

\end{table}

\begin{table}[h]
  \caption{Examples of Notebooks and their Metrics}
  \centering
  \label{tab:notebookmetrics}
    \begin{tabular}{||c|c|c|c|c|c|c|c|c|c||}
        \hline
        \rotatebox{90}{ID}&\rotatebox{90}{Lines of Code}&\rotatebox{90}{Lines of Comment}&\rotatebox{90}{Num. of Code Cells}&\rotatebox{90}{Num. of MD* Cells}&\rotatebox{90}{Num. of Executed Cells}&\rotatebox{90}{Num.of Visual Data}&\rotatebox{90}{Num.of Headlines}&\rotatebox{90}{Performance Tier}&\rotatebox{90}{level of CU}\\
        \hline
        A&239 &69 &34 &11 &16 &1 &1 &1 &normal\\
        \hline
        B&38 &0 &2 &8 &0 &0 & 0 &4 &normal\\
        \hline
        C&166 &10 &27 &1 &11&0 &0 &0 &normal\\
        \hline
        D&405 &6 &76 &21 &5&2 &5 &4 &normal\\
        \hline
        E&214 &13 &44 &28 &16&10 &12 &0 &good\\
        \hline
        F&191 &2 &48 &46 &24&8 &8 &2 &good\\
        \hline
        G&99 &15 &32 &19 &46 &15 &17 &2 &good\\
        \hline
        H&140 &12 &22 &38 &18&5 &15 &2 &good\\
        \hline
    \end{tabular}
\\
\footnotesize{{* MD: markdown}}
\end{table}

Furthermore, our objective is to identify notebook metrics that have a significant impact on CU. To achieve this, we have calculated a subset of code metrics expected to serve as an effective indicator of CU. \tabref{notebookmetrics} displays the calculated metrics for these notebooks. The process of choosing and calculating them, as well as how their impact on CU was identified, is described in \secref{metrics}.

Based on the expert opinions in \tabref{notebookvote} and the metrics provided in \tabref{notebookmetrics}, we can hastily conclude that there is a direct relationship between some notebook metrics and CU.

We conducted a statistical analysis using a linear regression model to examine the relationship between individual notebook metrics and CU. The analysis considered CU (based on expert scores) as the dependent variable and various notebook metrics as independent variables. Results, detailed in \tabref{regresults}, indicate that specific factors significantly influence CU.
Notably, the inclusion of visual data types (
$\beta$=0.9383, p=0.0007) and the use of structured headlines ($\beta$=0.9405,p=0.0006) were identified as the most significant classifier of CU. While the number of executed cells ($\beta$=0.6440,p=0.0845) showed a relatively strong correlation.

\begin{table}[h]
\centering
\caption{Linear Regression for Level of CU Classification}
\label{tab:regresults}
\begin{tabular}{|l|c|c|c|}
\hline
\textbf{Variable} & \textbf{Correlation Coefficient)} & \textbf{p-value} & \textbf{Standard Error} \\ \hline
Lines of Code          & 0.1531              & 0.7074        & 0.0007             \\ \hline
Lines of Comment       & 0.0427              & 0.9198        & 0.0056             \\ \hline
Num. of Code Cells      & 0.1271              & 0.7565        & 0.0030             \\ \hline
Num. of MD Cells        & 0.4532              & 0.2558        & 0.0130             \\ \hline
Num. of Executed Cells  & 0.6440              & 0.0845        & 0.0093             \\ \hline
Num. of Visual Data     & 0.9383              & 0.0007        & 0.0058             \\ \hline
Num. of Headlines       & 0.9405              & 0.0006        & 0.0041             \\ \hline
Performance Tier         & -0.2582             & 0.5366        & 0.1358             \\ \hline
\end{tabular}
\end{table}

It appears that certain code metrics may play a significant role in understanding the code within notebooks. While this is an initial hypothesis based on a few straightforward examples, in this paper, we aim to investigate this effect on a larger dataset of notebooks and more extensive code metrics.


\section{Metrics of Notebooks}
\label{sec:metrics}
Code metrics play a vital role in evaluating the effectiveness, comprehensibility, and quality of computational notebooks.

In our previous work~\cite{mostafavi2024DistilKaggle}, we introduced a dataset named DistilKaggle, which contains 542,051 Jupyter notebooks along with their code metrics. This dataset provides information from notebooks in the form of two separate dataframes: one for code cells and another for markdown cells. Additionally, the process of identifying code metrics in Jupyter notebooks and the methods used to extract this information are explained in detail. Here, we briefly describe the most important aspects related to this dataset while omitting specific details.
We delve into two distinct groups of metrics: \emph{Notebook-Based Metrics} and \emph{Script-Based Metrics}. \emph{Notebook-Based Metrics} focus on the properties and structure of the notebook itself, including organization, documentation, and visual representation. On the other hand, \emph{Script-Based Metrics} concentrate on the underlying code, examining factors such as complexity, modularity, and adherence to coding standards. Through the exploration and analysis of these metrics, our aim is to gain deeper insights into the strengths, weaknesses, and overall effectiveness of computational notebooks.

\subsection{Notebook-Based Metrics}

These metrics are detailed in \tabref{features}. Subsequently, we will elaborate on metrics whose names may not clearly indicate their conceptual underpinnings.

\begin{table}[t]
\centering
\begin{threeparttable}[b]
\centering
\caption{Jupyter Notebook Code Metrics~\cite{mostafavi2024DistilKaggle}}
\label{tab:features}
\begin{tabular}{|p{5.5cm}|p{2cm}|p{2.5cm}|}
\hline
\textbf{Feature} & \textbf{Abbreviation} & \textbf{Reference}\\
\hline
\hline
\multicolumn{3}{|c|}{\textit{\textbf{Script-Based Metrics}}} \\
\hline
\hline
\multicolumn{3}{|c|}{\textit{Basic Metrics}} \\
\hline
Lines of Code & LOC &~\cite{scalabrino2019automatically,buse2009learning,kasto2013measuring,wang2022documentation,lavazza2023empirical}\\
\hline
Number of Blank Lines of Code & BLC &~\cite{scalabrino2019automatically,buse2009learning,kasto2013measuring}\\
\hline
Lines of Comments & LOCom &~\cite{scalabrino2019automatically,buse2009learning,kasto2013measuring}\\
\hline
Number of Comment Words & CW &~\cite{kasto2013measuring}\\
\hline
Number of Statements & S &~\cite{kasto2013measuring,scalabrino2019automatically}\\
\hline
Number of Parameters & P &~\cite{kasto2013measuring,scalabrino2019automatically}\\
\hline
Number of User-Defined Functions & UDF &~\cite{grotov2022large}\\
\hline
\multicolumn{3}{|c|}{\textit{Complexity Metrics}}\\
\hline
Cyclomatic Complexity & CyC &~\cite{scalabrino2019automatically,buse2009learning,kasto2013measuring,grotov2022large,lavazza2023empirical}\\
\hline
Nested Block Depth & NBD &~\cite{kasto2013measuring,scalabrino2019automatically}\\
\hline
Kind of Line of Code Identifier Density & KLCID &~\cite{klemola2003cognitive}\\
\hline
\multicolumn{3}{|c|}{\textit{Halstead Metrics}}\\
\hline
Number of Operands & OPRND &~\cite{posnett2011simpler,kasto2013measuring,scalabrino2019automatically}\\
\hline
Number of Operators & OPRAT &~\cite{posnett2011simpler,kasto2013measuring,buse2009learning}\\
\hline
Number of Unique Operands & UOPRND &~\cite{posnett2011simpler,kasto2013measuring}\\
\hline
Number of Unique Operators & UOPRAT &~\cite{posnett2011simpler,kasto2013measuring}\\
\hline
\multicolumn{3}{|c|}{\textit{Readability Metrics}}\\
\hline
Average Line Length of Code & ALLC &~\cite{buse2009learning}\\
\hline
Number of Identifiers & ID &~\cite{buse2009learning}\\
\hline
Average Number of Identifiers & AID &~\cite{buse2009learning}\\
\hline
\hline
\multicolumn{3}{|c|}{\textbf{\textit{Notebook-Based Metrics}}}\\
\hline
\hline
Number of Code Cells & CC &~\cite{pimentel2019large,wang2022documentation}\\
\hline
Mean Number of Lines in Code Cells & MeanLCC &~\cite{kasto2013measuring}\\
\hline
Number of Imports & I &~\cite{pimentel2019large,kasto2013measuring}\\
\hline
Mean Number of Words in Markdown Cells & MeanWMC &~\cite{wang2022documentation}\\
\hline
Number of H1 tags in the Markdown & H1 &~\cite{pimentel2019large,venkatesh2021automated,venkatesh2023enhancing}\\
\hline
Number of H2 tags in the Markdown & H2 &~\cite{pimentel2019large,wang2020assessing,venkatesh2023enhancing}\\
\hline
Number of H3 tags in the Markdown & H3 &~\cite{pimentel2019large,wang2020assessing,venkatesh2023enhancing}\\
\hline
Number of Markdown Cells & MC &~\cite{wang2022documentation,pimentel2019large,liu2021haconvgnn}\\
\hline
Mean Number of Lines in Markdown Cells & MeanLMC &~\cite{grotov2022large}\\
\hline
Number of Markdown Words & MW &~\cite{kasto2013measuring,yang2021subtle}\\
\hline
Number of Lines in Markdown Cells & LMC &~\cite{scalabrino2019automatically,wang2022documentation}\\
\hline
Distribution of Markdown Cells & DMC &~\cite{pimentel2019large,wang2022documentation}\\
\hline
Distribution of Imports & DI &~\cite{pimentel2019large,wang2022documentation}\\
\hline
Performance Tier & PT &~\cite{alomar2021behind,scalabrino2019automatically}\\
\hline
External API Popularity & EAP &~\cite{scalabrino2019automatically,zhu2021restoring,dilhara2021understanding,mostafavi2024can}\\
\hline
Number of Visualization Data Type & VDT &~\cite{ono2021interactive,agrawal2022understanding}\\
\hline
Number of Executed Cells & EC &~\cite{pimentel2019large}\\
\hline
\end{tabular}
\end{threeparttable}
\end{table}

\begin{itemize}
    \item \textbf{Number of Visualization Data Types}: This metric reflects the total number of visual outputs displayed by executing each cell in the notebooks. It encompasses various types of visual outputs, including images, graphs, and plots.

    \item \textbf{Number of Executed Cells}: This attribute indicates the total number of cells that have been executed in a notebook, each accompanied by its respective execution order number.

    \item \textbf{H1, H2, and H3 Headlines}: These metrics indicate the number of headings present within the markdown cells, typically used to denote titles and subtitles.

    \item \textbf{Distribution of Markdown Cells}: This metric refers to the way these cells are spread out throughout the content, reflecting how often they appear and how substantial their content is in terms of word count. A higher degree of imbalance in this distribution suggests a concentration of markdown content in fewer cells, while a more balanced distribution indicates a more even spread of markdown content across multiple cells.

    \item \textbf{Performance Tier}: This feature is included in the metadata of the Kaggle repository and determines the level of expertise of notebook developers based on specific criteria, categorized into levels ranging from 0 to 4\footnote{https://www.kaggle.com/progression}. Detailed information on how Kaggle calculates the performance tier can be found in \appenref{pt_discuss}.

    \item \textbf{External API Popularity}: This metric assigns a score to each notebook based on the popularity of the APIs and libraries used within it~\cite{zhu2021restoring}. A higher score indicates that other developers frequently utilize the APIs featured in the notebook, which has been shown to correlate with improved CU in previous studies~\cite{scalabrino2019automatically}.

\end{itemize}

\subsection{Script-Based Metrics}
Script-based metrics include seven basic metrics, three complexity metrics, four Halstead metrics and three readability metrics as introduced in \tabref{features}. Some metrics that require further explanation will be mentioned here:

\begin{itemize}
    \item \textbf{Basic metrics}:
    The number of lines of code, number of blank lines of code, number of comment lines of code, number of comment words, number of statements, number of parameters, and number of user-defined functions.

    \item \textbf{Halstead Metrics}: Halstead metrics were first employed by Parker and Becker~\cite{parker2003measuring} to measure and compare the effectiveness of students' solutions in two different coding assessments, based on the premise that these metrics can be viewed as indicators of work done. Kasto~\cite{kasto2013measuring} later identified four of Halstead's metrics as imperative metrics: the number of operands (the total count of variables and constants in the code), the number of operators (the total count of actions performed by the code, such as arithmetic operations and control flow statements), the number of unique operands (the count of distinct variables and constants), and the number of unique operators (the count of distinct actions represented in the code).

    \item \textbf{Cyclomatic Complexity}: Cyclomatic Complexity assigns a numerical value to code complexity based on the code's control flow graph. It was first utilized by Mathias~\cite{mathias1999role} to investigate the underlying nature of code designed for studying the process of program comprehension.

    \item \textbf{KLCID}: The Kind of Line of Code Identifier Density (KLCID)~\cite{klemola2003cognitive} is a complexity metric that analyzes the cognitive complexity associated with program comprehension tasks. The KLCID is calculated by counting the lines of conceptually unique code and determining the density of identifiers.
    The computation of KLCID involves transforming code into a format that identifies lines with similar structures. Unique lines containing identifiers are counted, reflecting distinct patterns in the code. KLCID is obtained by dividing the number of identifiers (such as variables, function names, or constants) by the total number of lines, thereby assessing cognitive complexity. This metric recognizes that lines with a higher density of identifiers are more challenging to understand, highlighting the comprehension difficulties posed by the use of dense identifiers in code segments.

\end{itemize}


\section{Challenges and Research Directions}
\label{sec:problem}

In the context of modern software development, computational notebooks have emerged as a vital tool for data scientists and researchers, particularly those using the Python programming language~\cite{de2024bug}. However, the quality of code and documentation within these notebooks often varies significantly, leading to challenges in code understandability and maintainability. This study seeks to assess code quality in Jupyter Notebooks by detecting deficiencies in understandability using static analysis metrics and transformer-based language models, with static metrics serving as inputs to our model. Furthermore, we provide insights into which factors, as captured by our metrics, most significantly influence CU.

Given the challenges posed by the limited understandability of Jupyter notebooks and the deficiencies in existing methods for assessing CU within this context, this paper presents a pioneering methodology for quantifying CU in Jupyter notebooks. Furthermore, it identifies the primary notebook metrics that significantly impact CU. To clarify the problem, in this section, we raise three research questions. Subsequently, we present our proposed approach (\secref{approach}) and evaluate its effectiveness (\secref{evaluation}) in addressing these questions.

\textbf{RQ1: How accurately is it possible to provide a criterion for measuring code understandability of computational notebooks based on user comments?}

In numerous studies~\cite{sykes1983effect,buse2009learning,medeiros2018investigating,scalabrino2019automatically,scalabrino2019automatically}, expert opinion has been employed as a criterion to measure CU. Questionnaires have been utilized as tools to assess the comprehension of experts based on their opinions regarding specific code snippets.
Unlike traditional methods that primarily focus on code structure or static analysis, our study integrates user opinions as a core component in measuring code understandability. What distinguishes our approach is that we analyze comments and feedback directly shared by users within notebook repositories, providing an additional layer of insight. This inclusion captures a broad spectrum of real-world perceptions and interactions from a large user base, which is often absent in methods that depend solely on structural metrics or small-scale, survey-based feedback.
However, two critical challenges must be addressed regarding these comments. First, we need to identify which comments are directly relevant to CU. Second, we must assess the extent to which these comments have been acknowledged or validated by other users. Considering these factors, it is essential to establish a criterion that effectively measures a notebook's understandability. Evaluating whether this criterion can independently assess CU or serve as a complement to previously established criteria is crucial. By comparing the outcomes derived from this new criterion against a baseline, we can determine if this approach represents an improvement over existing methods, thus confirming its effectiveness and reliability.

\textbf{RQ2: Can a model be developed that effectively classify code understandability of notebooks based on the evaluation metrics established in this study?}

Previous studies, such as~\cite{scalabrino2019automatically,buse2009learning,lavazza2023empirical}, have focused on developing methods for automatically measuring CU. While these studies primarily utilized Java-based programs due to their influence on the programming community, they did not provide a specific model applicable to other programming languages.
In order to develop this model, two factors must be taken into consideration: the code metrics relevant to notebook CU, and a suitable criterion for assessing CU based on the findings from RQ1. Previous studies have utilized upvotes as a metric to gauge the CU of large-scale notebooks datasets.

Therefore, a model trained using this criterion can serve as a valuable benchmark for comparison against our own model. If our method demonstrably outperforms previous methods on evaluation parameters, it can serve as an initial step toward addressing RQ3 in a positive manner.

\textbf{RQ3: Which notebook metrics are associated with code understandability, and how significantly do they influence the classification of code understandability?}

After identifying the metrics related to CU and learning how to extract them (as described in \secref{metrics}), addressing this research question requires considering the following points:

\begin{itemize}
    \item How can we measure the impact of these metrics on notebook CU?
    \item Which metrics are the most effective in classifying notebooks based on CU?
    \item Which group of metrics has the greatest impact on improving CU classification?
\end{itemize}

By exploring these questions, we aim to identify script-based and notebook-based metrics that either enhance or hinder CU. Understanding the most influential individual and grouped metrics will provide valuable insights into crafting more comprehensible notebooks, ultimately promoting better programming practices.

\section{The Proposed Approach}
\label{sec:approach}
The proposed approach consists of two key components. The first component of the proposed approach involves developing a criterion to assess the understandability of notebook code in \secsref{commentCU}{notebookCU}. This criterion was effective in addressing RQ1.

\begin{figure*}[t]
    \centering
    \frame{\includegraphics[width=0.9\linewidth]{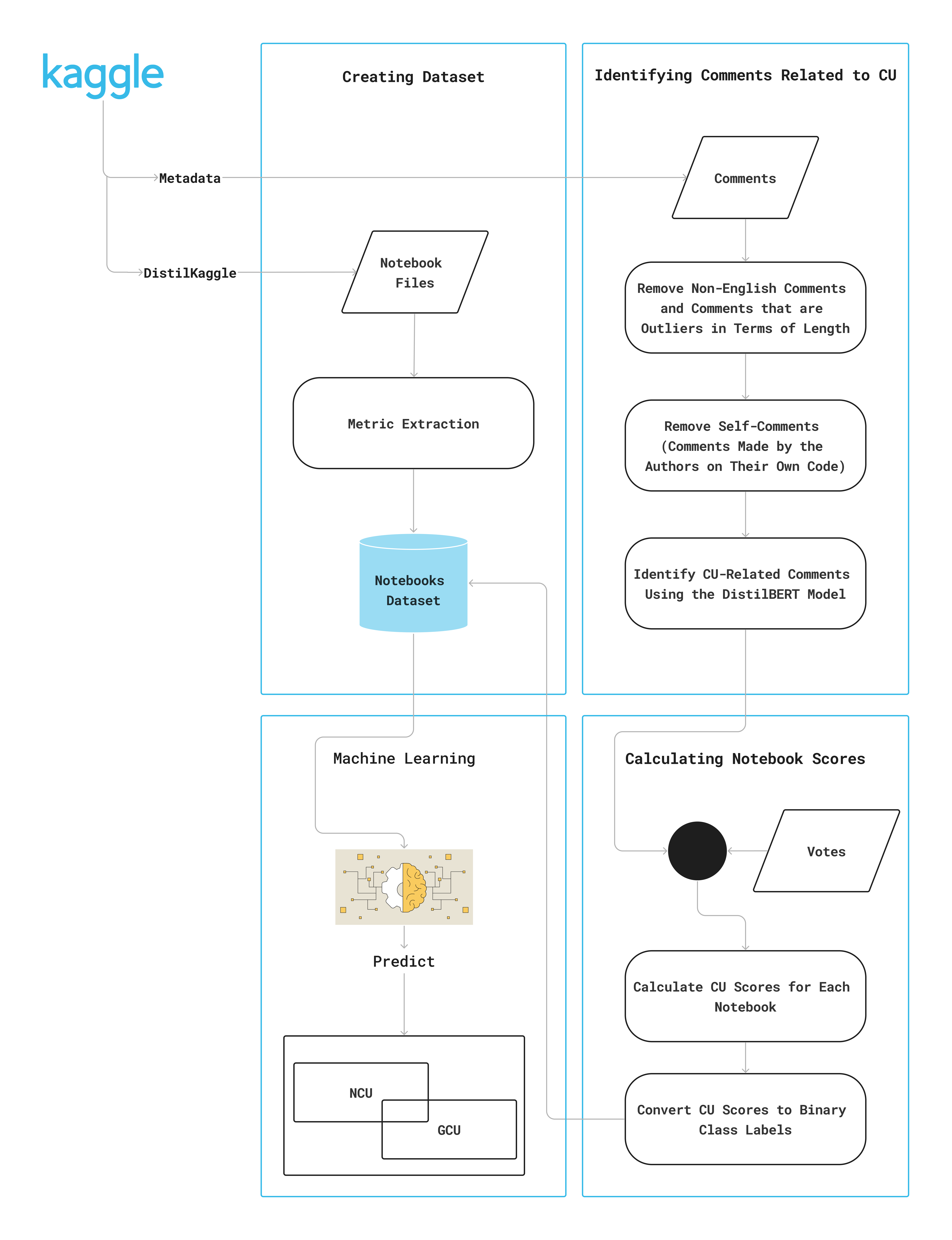}}
    \caption{Proposed Approach Overview to classify Two Class of Notebooks: Normal Code Understandability (NCU) and Good Code Understandability (GCU).}
    \label{fig:approach}
\end{figure*}

The second component involves preparing a comprehensive dataset of notebooks and generating the metrics shown in \tabref{features}. Different machine learning algorithms are then applied to find the best algorithm to determine the most effective metrics on CU. This analysis takes place in \secsref{minidataset}{MLO}.

In fact, in \secref{MLO}, we apply machine learning algorithms to address RQ2 and RQ3. These algorithms aim to identify relationships between input data and desired outcomes. The input data comes from the findings in \secref{minidataset}, while the desired outcomes are derived from the results in \secref{notebookCU}.

The purpose of the machine learning algorithms in \secref{MLO} is to learn how to accurately translate input data from \secref{minidataset} into the corresponding outcomes from \secref{notebookCU}. Once this learning process is complete, the algorithms can be used to classify notebooks based on new, unseen data. By discovering patterns between the notebook code metrics and their CU, these algorithms help provide insights into RQ2 and RQ3.



\figref{approach} outlines the four main steps of our approach to answer the research questions:

\begin{itemize}
    \item Identifying Comments Related to CU
    \item Calculating Notebook Score
    \item Preparing Dataset
    \item Utilizing Machine Learning
\end{itemize}

To start, \textit{Identifying Comments Related to CU} involves pinpointing comments that contribute to CU by reviewing the text annotations in code cells. This step filters out comments unrelated to comprehension, focusing specifically on those expressing positive or negative opinions about the code’s CU. These selected comments establish a baseline for assessing CU levels.
Next, \textit{Calculating Notebook Score} evaluates each notebook's overall quality and clarity. Scores are assigned based on multiple factors, including comment frequency, relevance to CU, and the number of views for each Jupyter notebook. These scores provide a quantitative assessment of each notebook’s understandability, facilitating effective comparisons across notebooks. The final scores also serve as labels for training machine learning models in subsequent steps.
These steps, along with \textit{Preparing Dataset} and \textit{Utilizing Machine Learning}, are discussed in detail in \secsref{commentCU}{MLO}.

\subsection{Identifying Comments Related to CU}
\label{sec:commentCU}

Numerous studies have leveraged user opinions for various software engineering activities. Lin et al.~\cite{lin2022opinion}, in their systematic review of opinion mining for software development, identified over 70 papers from leading research databases, like~\cite{panichella2015can,fu2013people,chen2014ar,kurtanovic2017mining}, that used user feedback to enhance software quality processes. These processes include evaluating software quality through crowd-sourced opinions, understanding general user satisfaction, assessing satisfaction with specific product features, identifying issues and feature requests from developer discussions and issue reports, and extracting actionable insights from user feedback.

In our study, we utilize user comments to detect CU in Jupyter notebooks. While user feedback may not always be wholly sufficient or perfectly consistent, it significantly complements expert analysis by providing a more comprehensive perspective on software understandability and user experience.
To achieve this, we used comments associated with each notebook in the DistilKaggle dataset~\cite{mostafavi2024DistilKaggle}, accessible through Kaggle metadata\footnote{https://www.kaggle.com/datasets/kaggle/meta-kaggle}. The large number of comments available in our dataset gave us the opportunity to conduct an exploratory study and analyze these comments with the aim of establishing a metric for Jupyter notebooks' CU.

During the data cleaning stage, any non-English comments were removed. We encountered a decline in performance during the fine-tuning of our deep learning models, primarily attributed to the presence of lengthy comments within our dataset. Consequently, by filtering out comments longer than the 99th percentile of the dataset's length distribution, we significantly reduced our fine-tuning time while retaining the majority of our data. These comments had between 1,000 and 17,000 characters, which is less than 1\% of the total data records.

Furthermore, comments posted by the authors of the notebooks themselves were excluded from the analysis. Similarly, comments that were shared in response to others' comments were also excluded. These types of responses often consist of simple expressions of gratitude or acknowledgments, which do not provide meaningful insights into the understandability of the notebook.

In the following, we outline our methodology for classifying user comments relating to Jupyter notebooks on Kaggle into three categories: ``positive in terms of understandability'', ``negative in terms of understandability'', or ``no relation to understandability'', utilizing a team of annotators. Initially, we crafted a guideline outlining definitions for each category supplemented with illustrative examples, as outlined in~\appenref{guidline1}. This guideline served to aid annotators in comprehending the process more effectively, thereby reducing discrepancies.
We had a total of 659,894 comments. Based on calculations similar to those outlined by Zhu et al.~\cite{zhu2021restoring}, we determined that a sample size of 1,039 comments was required to attain a confidence level of 99\% with a margin of error of ±4\%~\cite{sample2024}. Consequently, 1,039 comments were extracted to fine-tune our model.

The team were selected as annotators for the labeling procedure comprised four of the paper’s authors, each with a solid background in data science and experience working with Jupyter notebooks. These individuals were selected due to their familiarity with code comprehension challenges and their capacity to evaluate the clarity and readability of code effectively. The composition of the annotation team ensured a consistent and informed approach to labeling, leveraging their collective knowledge to identify comments relevant to CU accurately.

Each comment underwent assessment by two annotators who conducted an in-depth manual analysis of the content and assigned appropriate labels. An overall consensus was attained on 92\% of the labels, resulting in a Cohen’s kappa coefficient of 80\%, signifying reliable outcomes~\cite{cohen1960coefficient}. All contentious comments were meticulously reviewed, and a collaborative session was conducted with annotators to finalize the labeling decisions. Annotators engaged in productive discussions to harmonize and unanimously agree on the definitive labels.

After analyzing the labels, it was noted that there were six negative comments, which accounted for less than 1\% of the total comments. This pattern may have been influenced by the competitive nature among members participating in various data science competitions, ensuring that notebooks meet specific clarity and quality standards.

\begin{figure*}[t]
    \centering
    \frame{\includegraphics[width=0.5\linewidth]{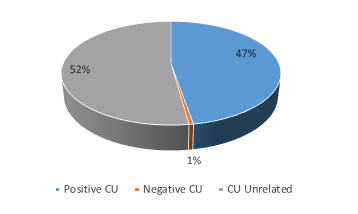}}
    \caption{Distribution of User Comments}
    \label{fig:comment_distribution}
\end{figure*}

The distribution of these three groups of comments is shown in \figref{comment_distribution}. For a better understanding of how to label comments, several examples of comments and their labels are given in  \appenref{comments_data}.

Considering this observation, we removed negative comments and consolidated the labels for each comment into two categories: ``related to code understandability'' (which are positive comments) and ``unrelated to code understandability''.
This distribution of comments motivates our paper's exploration of an intelligent approach for labeling notebooks as ``normal" or ``good" in terms of code understandability, as detailed in the \secref{notebookCU}.


Therefore, out of the total 994 results, 467 were categorized as ``related to code understandability" and 527 as ``unrelated to code understandability". To pinpoint comments specifically relevant to CU within the entire dataset, we fine-tuned the DistilBERT, a smaller general-purpose language representation model and a distilled version of BERT, which is smaller, faster, computationally cheaper, and lighter model on this data. DistilBERT was chosen due to its ability to effectively capture the contextual semantics of full sentences, which is crucial for understanding user comments. Additionally, as a distilled version of BERT, DistilBERT offers a good balance between performance and efficiency, making it suitable for our task. It is pre-trained on a large corpus of text, allowing it to understand the context of a sentence, and can be fine-tuned on a specific task, like classifying comments in this case.  We had two main objectives in fine-tuning this model. Firstly, we wanted to filter out comments that were unrelated to CU in order to evaluate the notebooks based only on their relevant comments. Secondly, we aimed to identify and include all comments related to CU when assigning an understandability score to the notebooks. We partitioned our dataset of $\sim1000$ comments into training, validation, and test sets using a split ratio of 70-20-10. The F1 score and accuracy were computed on the test split, ensuring an unbiased evaluation of the model's performance. The model achieved an F1-score of 90\% and \
accuracy of 92\%. We then applied this model to the entire dataset. The model identified 410,484 out of 659,894 comments in the whole dataset that were related to CU. Finally, a score of one was assigned to comments related to CU, while a score of zero was assigned to other comments.

\subsection{Calculating Notebook Score}
\label{sec:notebookCU}
To calculate the CU score of notebooks, we did two main tasks, which are presented in two sub-sections as follows.

\subsubsection{Calculating CU Scores}

As mentioned, the UOCU criterion is utilized to measure CU based on users' opinions. To calculate UOCU for each notebook, we use \eqaref{scoring} which involves a function \textit{f} that takes some metadata of a notebook and the BERTScore calculated in \secref{commentCU}.

\begin{equation}
    \label{eq:scoring}
    UOCU = [\sum_{i=1}^{n}(BERTScore_i * (CommentVote_i +1)) + \alpha] / \sqrt{TotalView}
\end{equation}

where:

- \textit{n} is the total comments of a notebook,

- \textit{BERTScore} is the binary score calculated for a notebook comment with DistilBERT,

- \textit{CommentVote} is the number of upvotes for each comment,

- \textit{TotalView} is all views of a notebook. \newline

Currently, DistilBERT's model generates a score for each comment that is either 1 (related to CU) or 0 (unrelated to CU). The details of \eqaref{scoring} are explained below.

  Apart from the relevance of a comment to the CU, the votes given to that comment show users' agreement. To augment this impact, we multiply it by ``(CommentVote + 1)''. Each comment starts with a value of one (+1), and with each additional upvote, this value increments by one.
   We calculated the total scores ($\Sigma$) of comments for each notebook, which yielded the initial notebook score.
   We divided the final notebook scores by the square root of the number of times the notebook was viewed on Kaggle. This was done to mitigate the impact of views on notebooks. This adjustment ensures that notebooks with higher view counts (and consequently more comments) do not receive disproportionately large scores, while notebooks with fewer view counts (and potentially fewer comments) do not receive unfairly lower scores.
   We have introduced a ``+$\alpha$ numerator" to differentiate notebooks with varying view counts that may not have CU-related comments. As illustrated in \eqaref{scoring}, excluding the ``+$\alpha$" would result in a numerator of zero for notebooks with zero comments, rendering them indistinguishable. With the inclusion of ``+$\alpha$" in the numerator, we generate a small value that is then divided by the square root of the total views, creating diverse numbers. Through iterative experimentation, we have consistently observed that a value of 5 yields superior results in accomplishing this objective.

\subsubsection{Converting CU Scores to Binary Class Labels}

Up to this point, we have effectively formulated the UOCU criterion leveraging DistilBERT. Should this criterion prove to be more fitting for comprehending notebook code compared to previous criteria (which is proved in the \secref{evaluation}), it can be adopted as the label for the dataset, enabling the training of a model to assess the understandability of notebooks.

In this section, our primary emphasis is on converting the continuous CU score into discrete binary labels. Although we assess machine learning performance using four-level and three-level class labels, the main focus remains on binary classification.

Continuous CU scores are less interpretable and cannot be directly used for classification. To address this, we can simplify the interpretation by discretizing. In common practice, variables are discretized into two categories (i.e., the predictor x falling above or below some threshold)~\cite{Andrew2009Splitting}. However, as shown in the study of Gelman et al.~\cite{Andrew2009Splitting}, we can do better by discretizing the objective variable into three values and throwing away the middle category. The study shows the efficiency of this method. It makes the general recommendation that the high and low categories each be set to contain 1/4 to 1/3 of the data, which results in comparisons with approximately 80\%–90\% of the efficiency of linear regression.

After finding the best scoring strategy and calculating the score for each notebook, we followed the same rule and discretized our score. We ranked notebooks according to their scores and divided them into four equally sized bins. Subsequently, we left out two bins in the middle and retained the first and the last quarters. The first quarter comprised notebooks with the lowest scores, labeled as Normal CU (NCU), and the last quarter encompassed notebooks with the highest scores, labeled as Good CU (GCU). Each of these selected bins contained 1/4 of the data.

\subsection{Dataset}
\label{sec:minidataset}
Various platforms offer large datasets of notebooks, providing a wealth of information about the notebooks themselves and their creators. Kaggle, a subsidiary of Google, is an online community specifically designed for data scientists and machine learning practitioners. Kaggle allows users to discover and publish datasets, explore and construct models within a web-based data science environment, collaborate with other professionals in the field, and participate in competitions to solve data science challenges. Similar to social media platforms, Kaggle incorporates features such as following a user and liking shared notebooks or datasets. The abundance of comments found under shared notebooks, which aid in CU, makes it valuable for employing opinion mining and extracting knowledge. Users have the ability to vote for a notebook or comment and endorse its content. Considering these attributes, we believe that any repository encompassing these features would be a suitable choice for implementing the approach outlined in this paper.

We initiated our study by utilizing the DistilKaggle dataset that was introduced in our prior research~\cite{mostafavi2024DistilKaggle}. This dataset offers a detailed insight into the content structure of 542,051 Jupyter notebooks, facilitating in-depth analysis of both code and markdown cells. Additionally, it included a notebook code metrics dataset that concentrated on identified code metrics of the notebooks. The methodology for constructing this tiny dataset is elaborated in \secref{metrics}.

\subsection{Utilizing Machine Learning}
\label{sec:MLO}
We now have the features dataset and UOCU (as labels) ready to run a machine learning operation. The goal of this part is to discover if there is a relationship between the metrics of notebooks and their CU. It is also possible to provide a model that can estimate the level of understandability of a notebook. As the dataset features are all numbers and are not that complex, we employed four machine learning models to discover the relationship. We employed Random Forest, CatBoost, Decision Tree, and XGBoost models for the classification of notebooks into binary classes. CatBoost was included because of its native support for categorical features without the need for extensive preprocessing, making it particularly suited to our dataset's structure. Additionally, it has demonstrated strong performance and reduced computational costs compared to other boosting methods. While XGBoost is known for its efficiency, scalability, and exceptional performance in structured/tabular data problems. It incorporates advanced features like regularization and tree pruning, making it highly effective in minimizing overfitting. Decision trees are interpretable and can capture complex relationships within the data. Finally, Random Forest was included due to its robustness against overfitting, strong performance with high-dimensional datasets, and ability to handle both classification and regression tasks effectively. Its ensemble nature leverages the power of multiple decision trees to improve accuracy and reduce variance. By experimenting with these diverse algorithms, we aimed to find the best-performing model for our specific task of notebook CU classification. Additionally, their simplicity allowed us to concentrate on the analysis of the features rather than intricate model architectures or complex hyperparameter tuning.
The results of the evaluation metrics and the accuracy of these algorithms are presented in \secref{bin_ml_eval}. The Random Forest algorithm ultimately stood out with an accuracy of 89\% and F1-score of 88\%. We conducted a grid search on a subset of the data to explore different hyperparameter configurations. We found that the default settings for these algorithms yielded competitive results while significantly reducing the computational overhead.


\section{Evaluations}
\label{sec:evaluation}
In this section, we will describe the experimental design that is needed to answer research questions. We will also evaluate our approach for generating notebook scores based on notebook comments on Kaggle. After ensuring the quality of notebook scores, we train our models based on these scores and evaluate their performance in different settings. The quantitative and qualitative results of these evaluations and their comparison with the baseline will be described in each section.

\subsection{Experiment Design}
\label{sec:experimentdesign}
To answer the questions of this research, we first set up the experiments. We trained our model on Google Colab, a cloud-based platform with a 12.7GB RAM, a 78GB Disk, a T4 GPU, and a Python 3 Engine. The important issues in the design of our experiments are described below.

\subsubsection{Sampling}
\label{sec:sample_dataset}
To streamline and optimize the study, a filter was applied to eliminate notebooks. In our initial exploration of the DistilKaggle dataset, we observed a wide range of notebook view counts, from just a few views up to thousands. We systematically evaluated the performance of our UOCU criterion on notebooks with varying view counts. Specifically, we randomly sampled notebooks from the dataset and divided them into bins based on their view counts (e.g., 0-100 views, 100-500 views, 500-1000 views, and so on). We then calculated UOCU for each bin and found that it results in zero values for most of them. We then manually inspected these bins to further investigate the findings. Our findings consistently showed that the UOCU approach struggled to produce scores for notebooks with fewer than 500 views. We hypothesize that this is due to the relatively low number of comments and user interactions associated with these notebooks. These notebooks have the potential to attract many more comments over time, but for now, we decided to apply a 500-view filter to the dataset. This filtering step resulted in a reduction in the dataset size, from 542,051 to 132,723 notebooks.

Based on sampling calculations similar to those outlined by Zhu et al.~\cite{zhu2021restoring}, we determined that a sample size of 1,050 notebooks is required to attain a confidence level of 99\% with a margin of error of ±4\%. Specifically, we applied the sample size formula provided by Creative Research Systems~\cite{sample2024}, which aligns with the methodological approach used by Zhu et al., while adjusting for the parameters of our study. Consequently, these 1,050 notebooks were extracted and excluded from the training set and preserved to establish ground truth.

\subsubsection{Ground Truth}
\label{sec:groundtruth}

We designed a human labeling task to classify computational notebooks into two categories: NCU and GCU. This task evaluates the CU scores introduced in our study, following the annotation framework outlined in~\appenref{guidline2}. Specifically, annotators assessed aspects such as logical structure, descriptive code and documentation, and error handling—key factors influencing code comprehensibility. These characteristics were derived from prior research to ensure alignment with established standards of code understandability~\cite{rule2018exploration,pimentel2019large,perkel2018jupyter}.

To establish a ground truth dataset, we adopted the labeling methodology of Guzman et al.\cite{guzman2015ensemble} on the dataset detailed in\secref{sample_dataset}. Initially, we provided annotators with a set of instructions defining each class, along with examples illustrating various characteristics. This guideline helped standardize the classification process and minimize disagreements.

We invited 55 experts to participate in the labeling task, receiving responses from 42. These experts had an average of five years of experience working with Python in scientific or industrial contexts. Notably, over 90\% were familiar with Jupyter notebooks, meeting the criteria necessary for accurate labeling.

Notebook samples were randomly assigned to experts, with each notebook assessed by three annotators to ensure reliability. To facilitate review while concealing metadata such as user comments, upvotes, and views, the notebooks were hosted on \textit{nbviewer.org}, an online tool for viewing Jupyter notebooks in a browser without requiring a local installation. Annotators conducted manual analyses and assigned labels accordingly. Disagreements were addressed through discussions, and a final consensus was reached, particularly for cases with missing labels.

To measure inter-rater agreement and ensure labeling reliability, we employed Fleiss’ kappa, which accommodates multiple raters. This was necessary as Cohen’s kappa applies only to two raters. More details on Fleiss’ kappa are provided in~\secref{EvalMetrics}.
Our evaluation yielded a Fleiss’ kappa coefficient of 82\%, indicating almost perfect agreement and confirming the reliability of our labeling process~\cite{landis1977application}. The final dataset consisted of 566 notebooks (54\%) labeled as NCU and 484 (46\%) as GCU. We used these 1,050 labeled notebooks as ground truth to manually validate our UOCU criterion and other baselines, with results reported in~\secref{CUEval}.

\subsubsection{Baselines}
\label{sec:baselines}
For our dataset, we needed to find models and criteria suitable for the specific task of CU classification of Jupyter notebooks.

As outlined in \secref{related_work}, our study aligns with prior work that measures CU based on human opinions, including subjective evaluations, surveys, and community-based metrics. However, existing approaches have notable limitations:
\begin{itemize}
    \item Limited Sample Size in Opinion-Based Studies:
    Studies like Oliveira et al.~\cite{oliveira2020evaluating}, Buse and Weimer~\cite{buse2009learning}, and Scalabrino et al.~\cite{scalabrino2019automatically} rely on subjective opinions from a limited number of participants, which may not generalize well across diverse datasets and coding environments.
    \item Challenges in Human-Labeled Training Data:
    While some research has leveraged developer opinions to train models, manual annotation of large datasets remains a bottleneck for scalability~\cite{medeiros2018investigating,lavazza2023empirical}.
    \item Reliance on User Votes as a Proxy for Quality:
    Several studies use metrics like StackOverflow upvotes~\cite{nasehi2012makes}, GitHub stars~\cite{lu2018internal}, and Kaggle notebook votes~\cite{liu2021haconvgnn} as proxies for code understandability. However, these votes may capture other factors beyond CU, such as popularity or model performance~\cite{wang2022documentation,Wang2021WhatMakes,mondal2023cell2doc}.
\end{itemize}

Given these limitations, we chose upvotes as a baseline due to their prevalence in repository-based evaluations of CU.

To accomplish this, we selected three different methods:

\begin{enumerate}
\item The upvote approach: We considered the number of upvotes on a notebook as a potential proxy measure for CU. As detailed in \secref{related_work}, certain studies that evaluate CU via user opinions have employed criteria such as the number of upvotes, likes, and stars to gauge this aspect within software repositories. Hence, we adopt this metric as a scoring mechanism for measuring CU within our dataset.

\item The CodeBERT~\cite{codebert} model fine-tuned on upvotes: In this method, we fine-tuned a transformer model on our dataset and used that as the baseline for our task.

\item The PLBART~\cite{ahmad2021unified} model fine-tuned on upvotes: Also in this approach, we fine-tuned another transformer model and used it as another baseline for our task.
\end{enumerate}

CodeBERT has been pre-trained in both programming language (PL) and natural language (NL). It learns representations that are useful for various NL-PL applications such as natural language code search and code documentation generation. PLBART (Pretrained LAnguage- and Vision-based BART) is a state-of-the-art deep learning model that extends BART to incorporate both language and vision inputs, enabling it to generate high-quality text in multiple languages while understanding visual information.
We employed the Hugging Face Transformers library for our work~\cite{huggingface2019transformers}. We did not have access to labeled data for fine-tuning these transformers. We only had 1,050 notebooks as our ground truth, which were used as our final test dataset to validate the performance of our methods. To address the scarcity of training data, we employed a straightforward approach of ranking notebooks based on their upvotes, considering it as a rudimentary metric for assessing CU.
This approach aligns with previous studies, which often consider highly-voted notebooks as a proxy for well-documented notebooks~\cite{Wang2021WhatMakes,mondal2023cell2doc}, or utilize the number of stars and forks as proxies for the quality of data science GitHub projects~\cite{morakot2023mining}).

Therefore, we sorted the notebooks according to their upvotes. Subsequently, we curated our dataset by selecting an equal number of notebooks from both ends of the ranking spectrum – notebooks with the lowest scores (indicating minimal understandability score) and those with the highest scores (indicating significant understandability score). The specific number of notebooks chosen for inclusion in the dataset was determined experimentally.

\begin{figure}[t]
    \centering
    \frame{\includegraphics[width=0.6\linewidth]{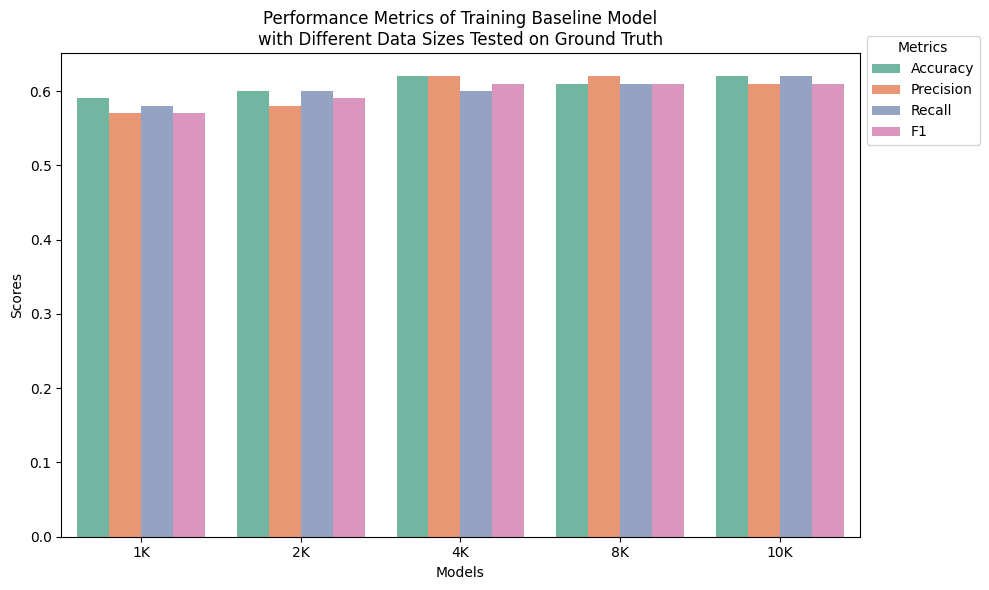}}
    \caption{The Impact of Changing Sample Size on Fine-Tuning Baseline Models}
    \label{fig:codebert_data_size}
\end{figure}

Our analysis, as depicted in \figref{codebert_data_size}, revealed a trend wherein increasing the sample size for fine-tuning baseline models led to improved performance, up to a threshold of 4,000 samples. Beyond this threshold, we observed no significant improvements in model training performance. Based on these findings, we concluded that a total sample size of 4,000 notebooks would be optimal for our study, comprising 2,000 notebooks with the lowest upvote scores, referred to as `NCU', and an additional 2,000 notebooks with the highest scores, referred to as `GCU'.

We partitioned our dataset of 4,000 samples into training, validation, and test sets using a split ratio of 70-20-10. To tokenize the input text, we employed both the CodeBERT and PLBART base tokenizers, both derived from the RoBERTa tokenizer architecture. Our tokenization process involved separate tokenization of markdown and code segments, followed by truncation of each part. In our investigations with CodeBERT, we identified 350 characters as the optimal token limit for markdown, allocating the remaining space to the code segment. Similarly, for PLBART, we assigned 512 characters to markdown and allocated the remaining space to the code segment.

Regarding training parameters, we employed accuracy metric to assess model performance. We configured the ``evaluation\_strategy" as ``epochs" and conducted training over 10 epochs. We restricted the number of epochs to 10, as a result of an observation that the model reached convergence on the input data within 10 epochs. We also set arguments regarding whether or not to run training on the training set, evaluation on the validation set, and classification on the test set all to ``True". Notably, our code was executed on a system with a GPU, as detailed in \secref{experimentdesign}.

\subsubsection{Evaluation Metrics}
\label{sec:EvalMetrics}
We utilize Accuracy, Precision, Recall, and F1-score to evaluate our classifiers. Accuracy calculates the percentage of all notebooks that are predicted correctly (as either GCU or NCU) out of all the existing notebooks in our ground truth. It is calculated as \eqaref{accuracy} where True Positives are the correctly predicted GCUs, and True Negatives are the correctly predicted NCUs. Precision, in the context of GCUs as positive labels, measures the proportion of correctly predicted GCU notebooks out of all notebooks predicted as GCUs. It is calculated as \eqaref{precision} where False Positives are the NCUs that were incorrectly predicted as GCUs. Recall, also known as Sensitivity or True Positive Rate, is the proportion of correctly predicted GCU notebooks out of all actual GCU notebooks in the dataset. It is calculated as \eqaref{recall} where False Negatives are the GCUs that were incorrectly predicted as NCUs. The F1-score is the harmonic mean of precision and recall. It provides a single metric that balances the trade-off between precision and recall, especially useful when dealing with class imbalance. It is calculated as \eqaref{f1-score}.

Another commonly employed metric in machine learning for evaluating binary classification is the AUC-ROC curve. It summarizes the classifier's ability to distinguish between the positive and negative classes across all possible threshold values. A higher AUC value indicates better discrimination power of the model. An AUC of 0.5 indicates random guessing, while an AUC of 1 indicates perfect classification.

In order to assess inter-rater agreement in our human labeling task among three raters, we use Fleiss' kappa~\cite{Fleiss1971Measuring}. Fleiss' kappa is a statistical measure used to assess the agreement among multiple raters when categorizing items into two or more categories. It is an extension of Cohen's kappa, which is designed for two raters.

In research, Fleiss' kappa is commonly interpreted similarly to Cohen's kappa. It ranges from 0 to 1, where:
\begin{itemize}
\item Values less than 0 indicate agreement worse than chance (very poor agreement).
\item 0 indicates agreement that is no better than chance (poor agreement).
\item Values between 0.01 and 0.20 indicate slight agreement.
\item Values between 0.21 and 0.40 indicate fair agreement.
\item Values between 0.41 and 0.60 indicate moderate agreement.
\item Values between 0.61 and 0.80 indicate substantial agreement.
\item Values between 0.81 and 1 indicate almost perfect agreement.
\item 1 indicates perfect agreement among raters.
\end{itemize}

\begin{equation}
    \label{eq:accuracy}
    \text{Accuracy} = \frac{\text{TP} + \text{TN}}{\text{TP} + \text{TN} + \text{FP} + \text{FN}}
\end{equation}

\begin{equation}
    \label{eq:precision}
    \text{Precision} = \frac{\text{TP}}{\text{TP} + \text{FP}}
\end{equation}

\begin{equation}
    \label{eq:recall}
    \text{Recall} = \frac{\text{TP}}{\text{TP} + \text{FN}}
\end{equation}

\begin{equation}
    \label{eq:f1-score}
    \text{F1-score} = 2 \times \frac{\text{Precision} \times \text{Recall}}{\text{Precision} + \text{Recall}}
\end{equation}

\subsubsection{Feature Importance Method in Machine Learning}
\label{sec:featue_imp}

Feature importance is a technique employed in data science to identify the most influential features or variables affecting a specific outcome. This information proves valuable when constructing predictive models or determining which features warrant deeper analysis. Various methods exist for calculating feature importance, such as permutation importance, mean decrease impurity, and SHAP (SHapley Additive exPlanations) values. Among the three methods, SHAP values are model agnostic, provide Interpretability for both global and local features, account for Feature Interactions, and are particularly useful for explaining the outcomes of machine learning models. In essence, they offer an estimation of each feature's contribution to the predicted outcome, considering interactions between features~\cite{LundbergL2017Unified,Scott2018Consistent}.

In the context of SHAP Beeswarm boxplots, each sample is represented by a point within each row of the plot. The horizontal placement of the point on the x-axis corresponds to the associated SHAP value. The density of points along each feature row reflects the strength of the relationship between the feature and the model's output. The range of colors, transitioning from blue to red, indicates the progression of metric values from minimum to maximum.

\subsection{Experiment Results}
After setting up the evaluation, in this section we will present the results after the experiment. After each experiment, we will describe the quantitative and qualitative results.

\subsubsection{Evaluation of the CU Criteria}
\label{sec:CUEval}
We first present the results of five different methods for measuring CU:

\begin{enumerate}
\item The upvote approach: In this approach, the quantity of upvotes bestowed upon a notebook is regarded as a metric for assessing CU. This method is detailed in \secref{baselines}.

\item The CodeBERT model fine-tuned on upvotes: In this method, we utilize the model explained in \secref{baselines}.

\item The PLBART model fine-tuned on upvotes: Also in this approach, we utilize the model explained in \secref{baselines}.

\item UOCU approach: As described in \secref{notebookCU}, this method uses users' opinions as a criterion for CU.

\item The hybrid approach:
In this approach, we address outliers, normalize the scores from Approach 1 and Approach 4, and then consider their sum as a new evaluation criterion. Outliers were identified based on the number of upvotes and the UOCU score. Specifically, approximately 1\% of the data with both low upvotes and low UOCU scores were removed. To normalize the scores from Approach 1 and Approach 4, we applied Min-Max normalization ~\cite{Aksu2019normalization} using the formula in \eqaref{min-max-normalization}:

\begin{equation}
    \label{eq:min-max-normalization}
    A' = \frac{A - \min(A)}{\max(A) - \min(A)}
\end{equation}

where \( A \) represents the original score, and \( A' \) is the normalized score. After normalization, the scores from both approaches were aggregated to create a hybrid score.

\end{enumerate}
\begin{table}[t]
  \caption{CU Score Evaluation}
  \centering
  \label{tab:CUEval}
  \begin{tabular}{|c c c c|}
    \hline
    &\textbf{Score}  & \textbf{F1-score}& \textbf{Accuracy} \\
    \hline
    1&\texttt{Upvote Approach} & 0.55 & 0.59\\
    \hline
    2&\texttt{CodeBERT fine-tuned by Upvote}& 0.61& 0.62\\
    \hline
    3&\texttt{PLBART fine-tuned by Upvote}& 0.65& 0.66\\
    \hline
    4&\textbf{UOCU Approach} & \textbf{0.66} & \textbf{0.70}\\
    \hline
    5&\textbf{Hybrid Approach} &  \textbf{0.86}& \textbf{0.87} \\
    \hline
  \end{tabular}
\end{table}

The outcomes generated from the five aforementioned methods are evaluated against the ground truth dataset (see \secref{groundtruth}), with detailed findings summarized in \tabref{CUEval}. As shown, the hybrid method achieves the highest Accuracy and F1-score. While the UOCU approach on its own demonstrates superior values in terms of F1-score and Accuracy compared to the previous methods, its amalgamation with users' upvotes yields even better results. Consequently, we incorporate the hybrid method as a label within our machine learning operations. 

\vspace{\baselineskip}

\begin{mdframed}
\textbf{RQ1}. \textit{How accurately is it possible to provide a criterion for measuring code understandability of computational notebooks based on their metadata?}
\par \noindent
\textbf{Answer:} Based on the results we achieved, it is possible to predict the CU scores of the notebooks on Kaggle to a great extent based on their metadata. With a hybrid approach based on the UOCU score and upvoted notebooks, we achieved an F1-score of 86\% and accuracy of 87\% in predicting CU category, which represents a significant improvement over other baselines.
\end{mdframed}

\subsubsection{Evaluating the Classifier}

Having evaluated our CU scores in the preceding section, we can now employ these scores as labels within our extensive dataset to facilitate the training and evaluation of our models.

Based on \secref{CUEval}, we found the best strategy for evaluation of CU, the hybrid approach. As mentioned in \secref{notebookCU}, we had calculated the hybrid CU scores in binary classification mode, and omitted the second and third quartiles (the middle two quartiles). As a result, a total of 66,360 (out of $\sim 132,720$) notebooks remained in the dataset, comprising 33,180 notebooks labeled as NCU and 33,180 notebooks labeled as GCU.
The range of scores defining each quartile and the corresponding number of notebooks are as shown in \tabref{cu_quartiles}.

\begin{table}[h]
    \centering
    \caption{Range of Hybrid CU Scores and The Number of Notebooks in Each Quartile}
    \label{tab:cu_quartiles}
    \begin{tabular}{|c|c|c|}
        \hline
        \textbf{Quartile} & \textbf{CU Score Range} & \textbf{Number of Notebooks} \\
        \hline
        Q1 & 0.01 – 0.06 &  33,180 \\
        Q2 & 0.06 – 0.08 &  33,180 \\
        Q3 & 0.08 – 0.11 &  33,180 \\
        Q4 & 0.11 – 1.84 &  33,180 \\
        \hline
    \end{tabular}
\end{table}


In preparation for training, we divided these notebooks into three distinct sets: the training set, the evaluation set, and the test set, with ratios of 70\%, 20\%, and 10\% respectively. In the training phase, we employed the evaluation set to mitigate overfitting and utilized the test set to optimize the hyperparameters of our candidate machine learning models. Ultimately, to determine the optimal model, we assessed the trained models using our dataset of 1,050 ground truth notebooks as introduced in \secref{groundtruth}. All numerical results presented for machine learning algorithm outcomes in binary classification are based on this ground truth dataset.

Before examining the accuracy of machine learning algorithms in binary classification, it is useful to first evaluate their accuracy across two different modes: four-level classification and three-level classification.

In our analysis, we employed three classification modes to evaluate the CU of notebooks:

\begin{itemize}
    \item Binary Classification: This mode categorizes notebooks into two classes: Good Code Understandability (GCU) and Not Code Understandable (NCU).

    \item Ternary Classification: In this mode, notebooks are divided into three equal parts based on their CU scores, allowing for a more nuanced understanding of their quality.

    \item Quaternary Classification: This classification reintroduces two additional classes that were excluded in the binary classification, resulting in four distinct categories for evaluation.

\end{itemize}

\tabref{four_ml_eval} presents the classification results for the quaternary classification of notebooks, detailing the accuracy and F1-score achieved by our machine learning models across the four defined categories.

\begin{table}[t]
    \centering
    \caption{Machine Learning Algorithm Results for Quaternary Classification}
    \begin{tabular}{|c c c c c|}
        \hline
         \textbf{Algorithm} & \textbf{Accuracy} & \textbf{Precision} & \textbf{Recall} & \textbf{F1-score} \\
        \hline
         {CatBoost} & 0.39 & 0.38 & 0.40 & 0.39 \\
        \hline
         {XGBoost} & 0.37 & 0.36 & 0.37 & 0.37 \\
        \hline
         {DecisionTree} & 0.30 & 0.30 & 0.30 & 0.30     \\
        \hline
         {Random Forest} & 0.38 & 0.36 & 0.39 & 0.36  \\
        \hline
    \end{tabular}
    \label{tab:four_ml_eval}
\end{table}

\tabref{tri_ml_eval} presents the evaluation metrics for ternary classification, illustrating the performance of the models in classifying notebooks into three equal classes.

\begin{table}[t]
    \centering
    \caption{Machine Learning Algorithm Results for Ternary Classification}
    \begin{tabular}{|c c c c c|}
        \hline
         \textbf{Algorithm} & \textbf{Accuracy} & \textbf{Precision} & \textbf{Recall} & \textbf{F1-score} \\
        \hline
         {CatBoost} &0.49  & 0.49 & 0.50 & 0.49 \\
        \hline
         {XGBoost} & 0.48 & 0.49 & 0.49 & 0.48 \\
        \hline
         {DecisionTree} & 0.40 & 0.41 & 0.41 & 0.41     \\
        \hline
         {Random Forest} & 0.49 & 0.48 & 0.50 & 0.49  \\
        \hline
    \end{tabular}
    \label{tab:tri_ml_eval}
\end{table}

Now that we've assessed the classification accuracy results for the three scenarios mentioned earlier, we can shift our focus back to evaluating the classification accuracy specifically for binary classification.

The classification results of the evaluation for our final dataset using four machine learning algorithms are presented in \tabref{bin_ml_eval}.

\begin{table}[t]
\centering
\caption{Machine Learning Results for Binary Classification (on hybrid approach)}
\begin{tabular}{|c c c c c c|}
\hline
 \textbf{Algorithm} & \textbf{Accuracy} & \textbf{Precision} & \textbf{Recall} & \textbf{F1-score} & \textbf{AUC-ROC} \\
\hline
 {CatBoost} & 0.79 & 0.77 & 0.77 & 0.77 & 0.84 \\
\hline
 {XGBoost} & 0.84 & 0.83 & 0.83 & 0.83 & 0.89 \\
\hline
 {DecisionTree} & 0.82 & 0.83 & 0.84 & 0.83 & 0.82   \\
\hline
 \textbf{Random Forest} & \textbf{0.89} & \textbf{0.88} & \textbf{0.89} & \textbf{0.88} & \textbf{0.94}  \\
\hline
\end{tabular}
\label{tab:bin_ml_eval}
\end{table}

As shown in \tabref{bin_ml_eval}, Random Forest produces the best results with an F1-score of 88\% and an Accuracy score of 89\%. We also utilize the AUC-ROC criterion, as introduced in \secref{EvalMetrics}, to provide additional validation for our experiments. Notably, this criterion yields a value of 94\% for the Random Forest algorithm,  reinforcing our confidence in its selection.

As far as we know, no prior studies have been conducted on measuring notebook CU using machine learning methods. Hence, we established a baseline relying on notebook upvotes and trained four machine learning models accordingly. The outcomes are presented in \tabref{bin_ml_base}. Notably, training based on the UOCU criterion has shown substantial advancements in accuracy and F1-score.

\begin{table}[t]
\centering
\caption{Baseline ML Results for Binary Classification (on upvotes)}
\begin{tabular}{|c c c c c|}
\hline
 \textbf{Algorithm} & \textbf{Accuracy} & \textbf{Precision} & \textbf{Recall} & \textbf{F1-score}  \\
\hline
 {CatBoost} & 0.76 & 0.71 & 0.74 & 0.74 \\
\hline
 {XGBoost} & 0.84 & 0.74 & 0.78 & 0.80 \\
\hline
 {DecisionTree} & 0.76 & 0.72 & 0.74 & 0.75 \\
\hline
 \textbf{Random Forest} & \textbf{0.82} & \textbf{0.75} & \textbf{0.79} & \textbf{0.80}  \\
\hline
\end{tabular}
\label{tab:bin_ml_base}
\end{table}

\vspace{\baselineskip}

\begin{mdframed}
\textbf{RQ2}. \textit{RQ2: Can a model be developed that effectively classify code understandability of notebooks based on the evaluation metrics established in this study?}
\par \noindent
\textbf{Answer:} We developed a Classifier using the Random Forest machine learning algorithm to evaluate CU in notebooks through binary classification. This endeavor yielded a commendable accuracy of 89\% along with an F1-score of 88\%, showcasing significant improvements in accuracy and F1-score compared to the baseline.

\end{mdframed}

\subsubsection{Finding Features Effective on CU}
\label{sec:bin_ml_eval}

\begin{figure}[t]
    \frame{\includegraphics[width=\linewidth]{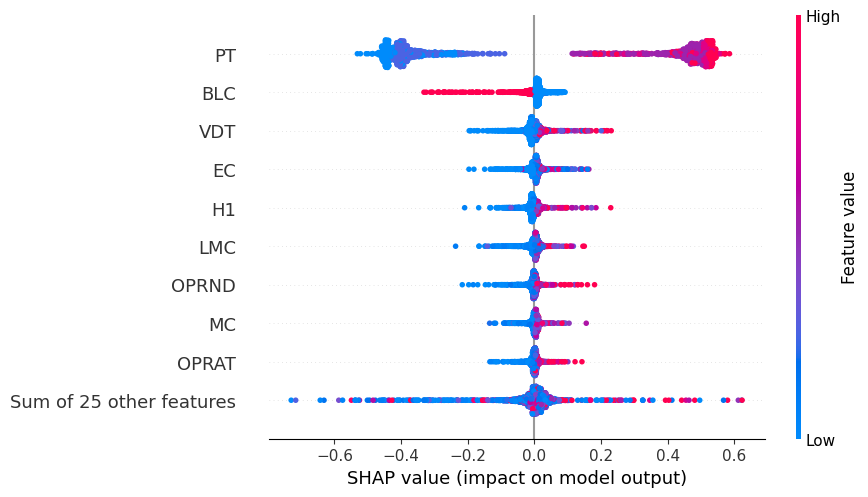}}
    \caption{The Beeswarm Boxplot for SHAP Values}
    \label{fig:shap with PT}
\end{figure}

In our study, we aimed to investigate the factors that impact CU in a dataset of notebooks. To achieve this, we applied the SHAP framework to assess the importance of different features, as discussed in \secref{featue_imp}. Our results are presented in \figref{shap with PT}, which illustrates a Beeswarm summary box plot. This plot offers a condensed overview of how the most significant notebook features impact the model's output.

\figref{shap with PT} of SHAP values reveal that  \textbf{performance tier (PT)} is the most important feature, on average, in determining CU. Users with lower performance tiers are generally less likely to produce understandable notebooks. Furthermore, the higher the expertise, the cleaner the notebook code. This result was consistent across our experiments with different learning algorithms and various measures of CU. The takeaway is straightforward: investing in training and enhancing the notebook developer experience is crucial for achieving a better understanding of the code.

Perhaps we are looking for a model capable of classification using static metrics extracted from a notebook, regardless of the developer's skill level. In this context, assigning a value to the PT metric might not be straightforward. In a separate experiment, we omitted this metric and attained an accuracy and F1-score of 80\% using the Random Forest algorithm.

According to the SHAP chart, the next most important features in determining CU are a script-based metric and four notebook-based metrics as follows:
\begin{itemize}
    \item \textbf{Number of blank lines of code (BLC)}, which is related to notebook code, has a negative impact on CU.
    \item \textbf{Visualization data types (VDT)}, which is related to code cell output, has a positive impact on CU.
    \item \textbf{Number of executed cells (EC)}, which is related to code cell output, has a positive impact on CU.
    \item \textbf{Number of H1 tags in markdown headlines (H1)}, which is related to markdown cells, has a positive impact on CU.
    \item \textbf{Number of lines of markdown cells (LMC)}, which is related to markdown cells, has a positive impact on CU.
\end{itemize}

\begin{table}[t]
    \centering
    \caption{The Importance of Each Group of Notebook Metrics in CU (on  hybrid  approach)}
    \begin{tabular}{|c c c c c c|}
        \hline
         \textbf{metric group} & \textbf{Best Algorithm} & \textbf{Precision}& \textbf{Recall} & \textbf{F1-score} & \textbf{Accuracy} \\
        \hline
        \hline
         \textbf{notebook-based} & Random Forest & 0.85 & 0.83 & 0.84 & 0.84 \\
        \hline
        \hline
         \textbf{script-based} & Random Forest & 0.77 & 0.76 & 0.76 & 0.76 \\
        \hline
        \hline
         \textit{basic metrics} & Decision Tree & 0.77 & 0.76 & 0.76 & 0.76 \\
        \hline
         \textit{complexity metrics} & Decision Tree & 0.75 & 0.74 & 0.74 & 0.75 \\
        \hline
        \textit{readability metrics} & Decision Tree & 0.75 & 0.76 & 0.75 & 0.76 \\
        \hline
        \textit{Halstead metrics} & Decision Tree & 0.77 & 0.76 & 0.76 & 0.76 \\
        \hline
    \end{tabular}
    \label{tab:metrics_impact}
\end{table}

We continued to investigate the impact of each group or subgroup of code metrics on code comprehensibility. The results of this investigation are provided in \tabref{metrics_impact}. This analysis aimed to determine which algorithm can best assess code understandability with the highest accuracy for each group of code metrics. The findings reveal that notebook-based metrics are more effective on their own compared to script-based metrics. Furthermore, within the category of script-based metrics, basic and Halstead metrics exhibit greater effectiveness.

\vspace{\baselineskip}

\begin{mdframed}
\textbf{RQ3}. \textit{RQ3: Which notebook metrics are associated with code understandability, and how significantly do they influence the classification of code understandability?}
\par \noindent
\textbf{Answer:}After training our models based on CU scores, we used Shapley scores to uncover the most impactful features of CU. At a broader level of analysis, notebook-based metrics exhibit the strongest influence on CU. Drilling down further, metrics like the number of blank lines of code, number of visualization data types, number of executed cells, number of H1 tags in markdown headlines, and number of lines of markdown cells all play a crucial role.
\end{mdframed}


\section{Discussion}
\label{sec:discussion}

This section summarizes the results, compares them with existing findings, discusses potential applications, and highlights the novelty of our approach.

\textit{Findings:} The study consists of two phases; The first phase introduced a method to assess CU in Jupyter notebooks using Kaggle metadata and user feedback. Five approaches were tested, including fine-tuned CodeBERT, PLBART, upvote-based models, the UOCU approach, and a hybrid model combining the latter two. The hybrid model performed best, achieving 87\% accuracy. The second phase analyzed which notebook metrics impact CU the most. Machine learning models were applied to CU scores, with SHAP analysis identifying key factors. Notebooks from proficient users were more understandable, while excessive blank lines reduced clarity. Metrics like visualization outputs, executed cells, top-level headers, and markdown lines positively influenced CU, highlighting the importance of clean code, documentation, and visual aids.
Additionally, notebook-based metrics (markdown quality, execution clarity, visualizations) had a stronger impact on CU than traditional script-based metrics, emphasizing the unique, interactive nature of computational notebooks.

\textit{Comparison to Existing Work}: Our findings underscore the performance tier (PT) as a significant factor influencing CU. This aligns with prior studies that suggest developer expertise impacts code quality, understandability or refactoring. For instance, Alomar et al.~\cite{alomar2021behind} show that experienced developers are more inclined to refactor, improving code quality and readability. Similarly, Scalabrino et al.\cite{scalabrino2019automatically} reveal that CU is often more affected by the developers' proficiency than by specific code snippets.


Additionally, the positive effect of metrics such as Header 1 and Lines of Markdown Cells on CU underscores the importance of documentation, particularly the use of markdown cells, in enhancing notebook CU. Prior studies support this finding, highlighting the role of documentation in improving readability and understandability. For example, Börstler et al.~\cite{borstler2023developers} examine how practitioners perceive and discuss various aspects of code quality, including understandability. Documentation is identified as a key factor influencing CU, with developers frequently associating well-documented code with higher quality.
Furthermore, a considerable number of previous studies have highlighted the positive impact of markdown on code understandability. However, unlike our study, the extent of this effect has not been quantitatively evaluated on a large dataset~\cite{pimentel2019large,venkatesh2021automated,dong2021splitting}.

Also, our study provides novel insights into how the number of executed cells and the visualization data types in Cell Output specifically affect CU in computational notebooks. These metrics, which are unique to the notebook environment, relate directly to the outputs generated by code cells. Our findings indicate that when the outputs are limited or when there is a scarcity of visual elements—such as images, plots, and diagrams—the understandability of the notebook decreases. This underscores the essential role of visualizations in enhancing code understandability, as they provide intuitive context that aids in interpreting results and understanding the flow of analysis.

\textit{Novelty and Contributions}: Our study provides a novel perspective by leveraging user feedback as a metric for measuring CU, an approach not previously explored in code understandability research. Unlike traditional methods that primarily focus on code structure or static analysis, our study integrates user opinions as a core component in assessing the CU of Jupyter notebooks—a widely-used format in data science and machine learning that has not been extensively explored in CU research. What distinguishes our approach is the analysis of comments and feedback directly shared by users within notebook repositories, adding an additional layer of insight. This inclusion captures a broad spectrum of real-world perceptions and interactions from a large user base, which is often absent in methods that rely solely on structural metrics or small-scale, survey-based feedback.

Furthermore, after determining the most effective method for measuring CU (the hybrid approach combining user opinions and notebook upvotes), we employed machine learning techniques to identify the notebook metrics that most significantly influence CU. By utilizing multiple machine learning algorithms, we presented, for the first time, a model capable of assessing notebook CU levels. This enabled us to identify the most impactful metrics, such as performance tier, markdown quality, and visualization elements, providing actionable insights for improving notebook comprehensibility.

\textit{Implications and Potential Applications}:
This section outlines the applications of our approach and the generalizability of our findings. By leveraging user feedback as a criterion for CU, this study highlights the potential for applying similar methods across various code repositories, platforms, and programming languages. Integrating user insights into CU assessments introduces a valuable context-sensitive criterion, creating a more holistic, user-centered understanding of code quality.

Our findings offer significant implications for notebook authors and tool developers aiming to enhance CU in Jupyter notebooks. Recognizing the influence of performance tier as a key factor suggests that resources like tutorials, training modules, and mentorship programs could help novice users develop coding expertise, thereby improving overall notebook quality. Additionally, notebook environments could introduce expertise-driven suggestions—such as reminders for more documentation or alerts about excessive blank lines—to support better readability, as our study found these aspects can detract from clarity.
Furthermore, the substantial impact of notebook-specific metrics, including markdown cells, headers, and visualizations, underscores the need for tools that encourage well-organized, interactive analyses. Enhancing applications like JupyterLab with extensions for visualization and markdown formatting can guide users in structuring their notebooks more effectively. This is particularly valuable in educational settings, where notebooks are widely used for teaching programming and data science. By promoting the use of rich visual content, such as plots, images, and interactive components, developers can help users create notebooks that are both functional and visually intuitive, ultimately facilitating a deeper understanding of complex analyses.

\section{Related Work}
\label{sec:related_work}

This section provides an overview of related work in the areas of ``Jupyter Notebooks Code Quality" and ``Measuring Code Understandability Based on Opinion".

\subsection{Jupyter Notebooks Code Quality}
High-quality code can expedite the process of code comprehension. Several studies related to code quality in Jupyter notebooks have been done in recent years.
In a qualitative study, Dong et al.~\cite{dong2021splitting} investigated how scientists clean their code on Jupyter notebooks. They sampled 20,000 notebooks with more than ten commits and classified them into three categories: personal, sharing, and productive notebooks. Finally, they sampled ten notebooks from each group and analyzed the changes between subsequent commits. Based on their findings, deleting commented blocks, adding commented blocks, reordering cells, adding markdown, renaming variables, splitting code into multiple Python files, and reorganizing code into functions and classes are among the most frequent activities done by data scientists to clean their code.

One of the activities that programmers engage in during the code comprehension phase of programming is executing different parts of the code and monitoring state variables. In a study by Wang et al.~\cite{wang2020assessing}, 10,000 GitHub repositories containing Jupyter notebooks were randomly cloned. They found that 73\% of the repositories were not reproducible, which can have a negative impact on CU.

Venkatesh et al. introduced an approach aimed at enhancing comprehension and navigation within Jupyter notebooks~\cite{venkatesh2023enhancing}. The paper addresses the challenges associated with undocumented notebooks. The authors propose a tool named HeaderGen to tackle these challenges. HeaderGen automatically supplements code cells with annotations derived from a taxonomy of machine learning operations. This augmentation aims to enhance both comprehension and navigation within notebooks. In a user study, participants reported finding HeaderGen beneficial.

In a study by Zhu et al.~\cite{zhu2021restoring}, over 4,000 Kaggle notebooks were sampled to investigate the causes of a large portion of repositories being unexecutable. The study found that the use of deprecated APIs was the primary cause of 31\% of unexecutabilities. To address this issue, they developed a library called Relancer that can automatically diagnose, infer, and upgrade deprecated APIs using a divide and conquer approach.

While the ability to execute code cells in different orders can help data scientists try out different settings and find the best one, it also presents a challenge to the reproducibility of results. The order of code cells in a notebook may not match the order in which the original developer executed them. To address this issue, Wang et al. developed a tool that can automatically find the possible orders that satisfy the dependencies between code cells~\cite{wang2020assessing}.

Wang et al.~\cite{wang2020better} used the PEP8 checker to analyze the code quality in Jupyter notebooks shared in two Kaggle competitions. The PEP8 checker validates whether Python code follows Python coding style best practices, such as line length and spacing between operators. They observed that 36.26\% of lines of code in Jupyter notebooks led to PEP8 checker errors, which may negatively affect CU. In contrast, when the PEP checker was applied to another dataset of Python scripts, only 13\% of Python lines were erroneous. Our study clearly demonstrates that the quality of Jupyter notebook code is significantly lower than that of regular Python scripts based on PEP checker metrics. Grotov et al.~\cite{grotov2022large} conducted a comparison between Python code written in Jupyter notebooks and traditional Python scripts, examining them from two perspectives: structural and stylistic. In terms of structure, they discovered several distinct aspects in which notebook code differs from its traditional counterpart. Notebooks exhibited a unique pattern in the utilization of functions, and the complexity metrics indicated that they contain code that is structurally simpler but more intertwined. This could potentially decrease their comprehensibility. In terms of stylistic differences, the disparities were more pronounced, with notebooks having 1.4 times more stylistic errors compared to scripts.

Literate programming is a programming paradigm that involves adding human-understandable natural language to small code snippets to make them as easy to understand as possible~\cite{knuth1984literate}. Pimentel et al.~\cite{pimentel2019large} used the GitHub API to analyze GitHub repositories between 2013 and 2018, and found that almost all Jupyter notebooks had markup cells, which is a characteristic of literate programming.

Wang et al.~\cite{wang2022documentation} compared a limited number of notebooks (100 notebooks) with high upvotes to a large dataset of randomly selected notebooks on Github. After observing that the notebooks with higher scores had an average number of markdown lines greater than the other group, they concluded that markdown cells and documentation play a significant role in the code quality of Jupyter notebooks. The authors then implemented a model named Themisto, which can automatically generate documentation for code cells. Like this study, our study also considers the number of documentation lines as one of the possible metrics that can affect code quality. However, unlike Wang et al.'s study, which was based on a limited number of notebooks, we evaluate this metric across a large number of notebooks in our dataset.

Brown et al.~\cite{brown2023facilitating} explore methods for facilitating dependency exploration in computational notebooks by focusing on tools and visualization techniques that help users understand cell dependencies, thereby enhancing reproducibility and comprehensibility of notebook workflows. Their approach primarily addresses the structural challenges of navigating inter-cell dependencies, offering visualization aids like dependency graphs and minimaps to simplify understanding.

Recent work by McNutt et al.~\cite{mcnutt2023design} explores how AI-powered code assistants can be optimized for computational notebook environments that have unique demands, which includes enhancing design features in support of iterative, exploratory coding. This points out the challenges in integrating AI-driven code suggestions into the typical nonlinear workflow inherent to notebooks and calls for meaningful customizability of design features.

Venkatesh et al.~\cite{venkatesh2023enhancing}  introduces a tool-based approach (HeaderGen) to improve notebook comprehensibility through static analysis and automated header generation, our work focuses on understanding and predicting CU by analyzing code metrics. HeaderGen primarily enhances navigation and structural coherence in machine learning notebooks by classifying functions and adding markdown headers, which aids in comprehension.

The studies mentioned attempted to measure code quality using various features derived from Jupyter notebooks, but none of them measured the extent to which each feature can affect CU. Some of these studies only analyzed a small number of notebooks, which may not be representative of the broader population of Jupyter notebooks on GitHub. There is currently no widely accepted set of metrics for evaluating Jupyter notebook code quality. Different studies may use different metrics, making it difficult to compare results across studies. Our study is different and complements these studies by applying a machine learning model to determine if each metric is can affect CU. We also report which features have a significant impact on code comprehension and which features have an insignificant effect.

\subsection{Measuring CU Based on Opinion}
In this section, we provide background information on studies related to our research, which aims to develop a comprehensive metric for understandability based on software developers' opinions.

Olivera et al.~\cite{oliveira2020evaluating} categorized the criteria used in a code comprehension study into five groups: correctness, time, opinion, visual metrics, and brain metrics. According to their findings, over half of the studies used people's opinions as a measure of CU. These studies relied on the subjective preferences and instincts of the participants instead of analyzing their actions or performance outcomes. In the following section, we will review some of these studies.

One such study is by Sykes et al.~\cite{sykes1983effect}, which investigated the level of programmers' understandability of the start-end states of blocks in Pascal programs. To achieve this, the study developed two types of questionnaires - a subjective questionnaire that asked people to express their understanding of the program in one sentence, and a code comprehension test with ten questions from different parts of the Pascal code.

Buse and Weimer~\cite{buse2009learning} explored the concept of code readability and investigated its relation to software quality. They collected data from 120 human annotators to derive associations between a simple set of local code features and human notions of readability. The annotators rated Java code snippets on a scale of one to five, where a higher score represented greater readability.

Medeiros et al.~\cite{medeiros2018investigating} aimed to investigate the relevance of misunderstanding patterns in C code by analyzing 50 open-source C projects on GitHub. They used a combination of repository mining and a survey to identify common misunderstanding patterns and evaluate their impact on understanding the source code. The survey presented several code patterns to participants and asked them about the level of influence each pattern had on their understandability of the code.

Scalabrino et al.~\cite{scalabrino2019automatically} conducted an extensive evaluation of 121 existing, as well as new, code-related, documentation-related, and developer-related metrics. They attempted to correlate each metric with understandability and build models that combine metrics to assess understandability. To do this, they used 444 human evaluations from 63 developers. One of the criteria they used to measure CU is called Perceived Binary Understandability (PBU), which is based on developers' opinions. This is a binary categorical variable that is true if a developer perceives that they understood a given code, and false otherwise. Finally, they obtained a bold negative result: none of the 121 experimented metrics is able to capture CU. As in the previous article, the methodology of our study is based on collecting opinions from a limited number of developers. The research scope of this article was limited to Java-based software.

Lavazza et al.~\cite{lavazza2023empirical} examined software understandability using metrics like Lines of Code, Cyclomatic Complexity, and Cognitive Complexity\footnote{Cognitive Complexity measures how difficult code is to understand by evaluating its control flow and nesting. It penalizes deep nesting and complex logic while rewarding readability-friendly constructs, focusing on how humans perceive code.}~\cite{campbell2018cognitive}. Their findings suggest that Cognitive Complexity does not significantly outperform traditional metrics in predicting CU, emphasizing the role of human factors alongside structural characteristics. They measured understandability through task completion times and applied machine learning to explore correlations between code measures and CU.

The challenges of these studies are that they relied on collecting opinions from a limited number of developers or on human labeling for the training dataset, respectively. Collecting opinions from a limited number of developers could limit the generalizability of the findings, as the opinions may not be representative of a broader population of developers. On the other hand, relying entirely on human labeling for the training dataset could limit the scalability of the approach, as it is not feasible to manually label large amounts of code.

There are some studies on software repositories that assess the quality and understandability of the code using factors like user ratings or votes. For instance, Nasehi et al.~\cite{nasehi2012makes} analyzed a large dataset of code snippets from StackOverflow and identified several features of high-quality code examples based on various evaluation metrics, including the number of upvotes, the number of views, code complexity, and the presence of error handling. Based on their analysis, they concluded that user feedback, as reflected by upvotes and downvotes, is a significant factor in determining the quality of code examples on StackOverflow.

Lu et al.~\cite{lu2018internal} consider user stars as a measure for identifying high-quality code in GitHub, although it was not the main focus of the study. The authors mentioned that stars are an indication of a project's popularity, and that a high number of stars can attract more developers to contribute to a project.

According to some researches~\cite{wang2022documentation,Wang2021WhatMakes,mondal2023cell2doc}, community voting is a good indicator of a computational notebook's quality. When community members upvote a notebook on Kaggle, they believe they are voting on the readability and completeness of the computational narrative. Liu et al.~\cite{liu2021haconvgnn} built their dataset around this hypothesis. They categorized notebooks with more votes on Kaggle as notebook with better code quality and well documentation.

The main challenge posed by the approaches presented in these studies is that they solely rely on user votes on software repositories. These metrics may be affected by factors beyond code comprehensibility, including ease of use, algorithm accuracy, popularity, and marketing considerations.

Despite the growing adoption of computational notebooks in data science and research, existing studies have primarily focused on notebook quality, reproducibility, and usability, without specifically addressing automated code understandability (CU) classification using notebook metrics. To the best of our knowledge, no prior work has systematically examined CU classification using a comprehensive set of 34 notebook metrics as classification features. Furthermore, no existing study has introduced an automated labeling approach like UOCU to assess CU. While some studies have explored aspects of notebook quality, these works have not proposed a structured methodology for CU classification, nor have they developed datasets and models tailored for this task. Our study fills this gap by presenting a novel approach that leverages notebook metrics for CU classification, providing a foundation for further exploration in this domain.

\section{Threats to Validity}
\label{sec:threads}
This section details potential threats to the validity and limitations of our findings, alongside our approaches to mitigate them.

\subsection{Internal Validity}
Internal validity refers to the degree to which any external factors or unanticipated biases may have impacted the design and analysis of the study. This includes confounding variables that could have affected the results~\cite{easterbrook2008selecting}.

A primary concern regarding our research's validity is the feasibility of determining the CU of notebooks based on user comments. This research hinges on our capability to automatically tag each notebook with an understandability label derived from these comments. To validate this approach, we consulted four experts, asking them to annotate the comments in relation to CU. The analysis revealed that a significant portion of comments indeed align with CU metrics. Furthermore, to determine if machine learning models could discern this relationship, we trained a model to predict the relevance of comments to CU; the model demonstrated promising results on test data.

It is a limitation of our approach that it only considers notebooks with more than 500 views for model fine-tuning. This threshold ensures a sufficient number of user comments for analysis but may introduce selection bias by excluding notebooks with fewer views. Consequently, our findings may not fully generalize to less-viewed notebooks, which could exhibit different understandability characteristics.

\subsection{External Validity}
External validity refers to how well the research questions align with the research objectives and the degree to which conclusions can be applied beyond the specific study~\cite{easterbrook2008selecting}.

While much of the existing research has evaluated CU primarily on the basis of code features, we identified that in Jupyter notebooks, the accompanying documentation plays a pivotal role in influencing CU. Recognizing this, we augmented our feature set, extending beyond the sole focus on code-related attributes, to include documentation-related aspects as well. As a result, our findings hold relevance not just for traditional coding environments such as Java or C, but notably for environments like Jupyter notebooks where markdown cells and documentation significantly bolster CU. This is further exemplified by how code comments, even within brief code snippets, can mirror the functionality of markdown cells in elevating understandability.

Even though we expanded the code criteria used by similar previous studies, we bounded our analysis to primitive criteria affecting CU. Motivated by low code quality in Jupyter notebooks and considering their important role for educational purposes, as the first study investigating CU in Jupyter notebooks, our primary goal was to take the first step toward devising a comprehensive metric for assessing CU in Jupyter notebooks. So, we bound our study to primitive code criteria and left the rest to future studies.

While our study focused on the metrics and features that could be feasibly derived from the Kaggle platform, it provides a concrete example for future studies to build upon and further extend these criteria.

Another limitation of our study is that our classification model categorizes notebooks into only two groups: normal and good CU. While this simplifies the analysis and ensures robust classification, it does not account for notebooks with poor CU or provide a more fine-grained ranking. Future work could explore multi-class or regression-based approaches to provide a more nuanced assessment of CU.

\subsection{Construct Validity}

Construct validity refers to whether the theoretical constructs in the study are accurately measured~\cite{easterbrook2008selecting}. In our study, a potential threat to construct validity is the use of upvotes as a proxy for measuring CU. While upvotes have been used in previous studies as proxies for code quality and user satisfaction, they may not always reflect true understandability, as they can be influenced by factors like the notebook’s popularity or visibility.

To address this, we incorporated not only upvotes but also user comments and view counts into our analysis. By factoring in user comments, we ensured that feedback related to understandability was directly considered. Additionally, by filtering out notebooks with high view counts but relatively low upvotes, we reduced the risk of including notebooks that gained upvotes merely due to exposure rather than understandability. This multifaceted approach helps mitigate the limitations of upvotes as a sole measure of CU.

Furthermore, our large dataset and the use of additional baselines, such as markdown usage and code complexity, contribute to a more comprehensive and reliable assessment of CU.
\subsection{Replicability}
We have made the experimental setup, data, and source code fully accessible to ensure the replicability of our results. The complete source code for our experiments, is publicly available at https://github.com/ISE-Research/NotebookCU. This repository provides detailed instructions for setting up the environment and reproducing the outcomes presented in this paper, enabling fellow researchers to validate and extend our findings. Additionally, our code, dataset, and models are available at https://doi.org/10.5281/zenodo.14827719.




\section{Conclusions and Future Work}
\label{sec:conclusion}

In this paper, we have introduced a novel metric for evaluating CU in computational notebooks, which leverages user comments. Our proposed methodology integrates the DistilBERT model to identify pertinent comments, incorporating upvotes and visits to determine CU. Notably, our measure has exhibited superior performance compared to four baseline models, as validated through a survey conducted by 42 experts who assessed 1,050 notebooks. Additionally, we gathered 34 metrics from a dataset of 132,723 notebooks optimized for machine learning operations, achieving an 89\% accuracy and an 88\% F1-score in predicting the impact of these metrics on CU via binary classification with the Random Forest classifier. SHAP charts revealed the most influential metrics to be the ``developer performance tier", ``number of blank lines of code", ``number of visualization data types", ``number of executed cells", ``number of H1 tags in markdown headlines", and ``number of lines of markdown cells", all playing a crucial role on CU.
As a matter of fact, while upvotes, traditionally utilized in prior studies for assessing CU, remain valuable, our results highlight that incorporating user comments and view counts can enhance the accuracy of this assessment. Consequently, based on our criterion, notebooks can be effectively classified with heightened accuracy utilizing the introduced notebook and code metrics.



\paragraph{\textbf{COMPLIANCE WITH ETHICAL STANDARDS}}

\paragraph{\textbf{Funding}} This research received no external funding.

\paragraph{\textbf{Ethical Approval}} Not applicable. This study does not involve human participants or animal research requiring ethical approval.

\paragraph{\textbf{Informed Consent}} Not applicable. No human subjects were directly involved in experiments requiring informed consent.

\paragraph{\textbf{Data Availability Statement}} Our dataset and experimental source code are available online at https://github.com/ISE-Research/NotebookCU.

\paragraph{\textbf{Conflict of Interest}} The authors declared that they have no conflict of interest.


\clearpage
\bibliographystyle{unsrt}
\bibliography{references}
\clearpage 
\begin{appendices}  

\section{- Performance Tier Calculation}
\label{appendix:pt_discuss}
The performance tier in Kaggle is divided into five levels, each reflecting different levels of expertise and achievement on the platform. Here are more details about each performance tier:
\begin{itemize}
    \item \textbf{Novice}: The Novice tier is the starting point for new users. Users typically start at this level when they join Kaggle and have not yet participated in competitions or completed tasks to earn points. Novice users are encouraged to explore the platform, learn, and gradually progress to higher tiers.
    \item \textbf{Contributor}: As users actively participate in Kaggle competitions and tasks, earn points, and contribute to the community, they progress to the Contributor tier. Contributors demonstrate a commitment to improving their skills, engaging with the Kaggle community, and making meaningful contributions to various projects.
    \item \textbf{Expert}: The Expert tier represents a significant milestone in a user's Kaggle journey. Users in the Expert tier have achieved a high level of skill and experience in data science and machine learning. They have demonstrated consistent success in competitions, gained recognition for their work, and actively share their knowledge with others in the community.
    \item \textbf{Master}: Users who reach the Master tier are among the top performers on Kaggle. Masters have a proven track record of success in competitions, often winning or ranking highly in challenging tasks. They are respected for their expertise, innovation, and ability to solve complex data problems effectively. Being a Master signifies a high level of proficiency and accomplishment in data science.
    \item \textbf{Grandmaster}: The Grandmaster tier is the highest level of achievement on Kaggle. Grandmasters are elite data scientists who have reached the pinnacle of success on the platform. They are highly skilled, innovative, and influential in the data science community. Grandmasters are admired for their exceptional performance, groundbreaking work, and leadership in the field.
\end{itemize}

Each tier represents a progression in a user's skills, knowledge, and impact within the Kaggle community. Advancing through the tiers requires dedication, hard work, continuous learning, and a passion for data science.

\section{- Annotation Guidelines for Classifying User Comments}
\label{appendix:guidline1}
Annotators should classify each comment according to its relevance to the understandability of the notebook. The comments fall into one of three categories:

\begin{enumerate}
    \item \textbf{Positive in Terms of Understandability}
    A comment should be classified as positive if it expresses appreciation, praise, or specific feedback regarding the clarity, ease of understanding, or overall readability of the notebook. Positive comments often contain words like ``clear'', ``understandable'', ``well-explained'' or ``easy to follow''.

    \textbf{Examples:}
    \begin{itemize}
        \item \textit{``Great job explaining the steps; it made the analysis very easy to understand!''}
        \item \textit{``This notebook is incredibly clear and easy to follow. Thanks for sharing!''}
    \end{itemize}

    \item \textbf{Negative in Terms of Understandability}
    A comment should be classified as negative if it highlights issues with the notebook's clarity, difficulty in understanding, or areas where explanations are confusing or lacking. Negative comments may include phrases such as ``hard to understand'', ``confusing'', or ``lacks explanation''.

    \textbf{Examples:}
    \begin{itemize}
        \item \textit{``It's difficult to follow the logic in this notebook.''}
        \item \textit{``The code is unclear, and there's not enough explanation.''}
    \end{itemize}
    \item \textbf{No Relation to Understandability}
    A comment should be classified as no relation to understandability if it does not directly address the notebook’s clarity or ease of understanding. This category includes comments that are unrelated to understandability, such as compliments on code efficiency, suggestions for additional features, general appreciation, or social interactions.

    \textbf{Examples:}
    \begin{itemize}
        \item \textit{``Good job on the data preprocessing steps!''}
        \item \textit{``Thanks for sharing!''}
        \item \textit{``Could you add more visualizations?''}
    \end{itemize}

\end{enumerate}

\section{- Distribution of Comments}  
\label{appendix:comments_data}

\begin{table}[h!]
    \centering
    \caption{A Sample of Comments Labeled with Experts}
    \label{tab:comment_tags}
    \begin{tabular}{|p{1.5cm}|p{5cm}|p{1.2cm}|p{1.4cm}|p{1.2cm}|}
        \hline
        \textbf{Kaggle\_Id} & \textbf{Comment} & \textbf{Positive CU} & \textbf{CU-unrelated} & \textbf{Negative CU} \\
        \hline
        388391 & Great score! & & 1 &  \\
        1296638 & Great work! Could you please checkout my work too. Thanks! \newline https://www.kaggle.com/aditya... & & 1 &  \\
        1165527 & One of the best notebook resources in Kaggle, appreciated. Upvoted. & 1 & &  \\
        654526 & thanks\textasciitilde & & 1 &  \\
        724117 & Excellent Notebook with interactive visualizations. Sometimes algorithms and ML are so funny. That's sarcasm. I'm 55 in Notebook and Dataset 254. Both are great. & 1 & &  \\
        1574460 & You changed my view toward data cleaning and also analyzing data. Cool Notebook! & 1 & &  \\
        1430133 & Thanks kaggle! I wanted to start working with kaggle for months. This was a pretty nice way to start and finally I did it :)) & & 1 &  \\
        3181480 & I don't understand the point of sharing a blend kernel, if you're not going to explain what you're blending. All I see is that you have a black box which produces 1.448 LB score & & & 1 \\
        1327863 & Good job @artaxae! This notebook is very informative. You have explained all the data processes in a detailed way. & 1 & &  \\
        310390 & thank you & & 1 &  \\
        1734304 & clean notebook! Thanks for sharing & 1 & &  \\
        1596760 & Nice example to get started in the competition & 1 & &  \\
        885042 & Well explained Ashok & 1 & &  \\
        1436169 & Nice work! I have a similar work based on emotion recognition from tweets. I would like to have your feedback on this notebook. & & 1 &  \\
        921444 & Nice Work. Notebook was very simple to follow !!! & 1 & &  \\
        605630 & Thank you for sharing! I learned so many things through this kernel. & 1 & &  \\
        1023754 & thank you for your work & & 1 &  \\
        577715 & Thanks for the sharing, learn a lot:) & 1 & &  \\
        1320208 & Thanks for sharing. I am a beginner in Python. These step by step explanations are very much useful for me. & 1 & &  \\
        825456 & wow, congrat! & & 1 &  \\
        1454626 & It was my first time using Random Forest! So much to learn & 1 & &  \\
        882902 & Really helpful notebook to learn ML, thanks for sharing your knowledge. & 1 & &  \\
        \hline
    \end{tabular}
\end{table}
\tabref{comment_tags} contains comments from Kaggle users along with labels indicating the impact of these comments on Code Understandability):
\begin{itemize}
    \item \textbf{Kaggle\_Id}: Unique identifier for a Kaggle comment written by the user.
    \item \textbf{Comment}: The text of the comment provided by the user.
    \item \textbf{Positive CU}: Indicates whether the comment positively contributes to the understanding of the code. A '1' means the comment is seen as helpful for improving code understandability; otherwise, it is left blank.
    \item \textbf{CU-unrelated}: Indicates whether the comment is unrelated to the code understandability. A '1' means the comment does not pertain to the code understandability; otherwise, it is left blank.
    \item \textbf{Negative CU}: Indicates whether the comment negatively impacts the understanding of the code. A '1' means the comment is viewed as detrimental to code understandability; otherwise, it is left blank.
\end{itemize}

\section{- Human Labeling Guideline for Notebook Classification}
\label{appendix:guidline2}

\paragraph{\textbf{Purpose}}

The goal of this human labeling task is to categorize notebooks into two categories: ``Normal Code Understandability'' (NCU) and ``Good Code Understandability'' (GCU) (Our initial observation on user comments cause to remove any label related to ``Weak Code Understandability''). This task is designed to assess the performance of the CU scores introduced in our study.

\paragraph{\textbf{Definitions}}
\begin{itemize}
\item \textbf{Code Understandability (CU)}: Code Understandability, also referred to as program understandability or code comprehensibility, is the active process by which developers acquire knowledge about a software system through exploring software artifacts, reading source code, and consulting documentation~\cite{xia2017measuring}.

\item \textbf{Normal Code Understandability (NCU)}: Notebooks labeled as NCU are functional and meet basic standards of readability but lack advanced organizational features, detailed explanations, or may require additional effort to fully grasp the workflow. They may contain redundant code, inconsistent formatting, or limited documentation.

\item \textbf{Good Code Understandability (GCU)}: Notebooks classified as GCU exhibit exceptional clarity, logical flow, thorough documentation, well-structured code cells, and effective use of visualizations. These notebooks require little effort for one to comprehend.
\end{itemize}
\paragraph{\textbf{Characteristics to Consider}}
According to scientific sources, the normal conditions for understanding notebook code include the following characteristics:

\begin{enumerate}
    \item \textbf{Logical Structure and Flow}: Ensures that the content is organized sequentially and coherently~\cite{rule2018exploration}.
    \item \textbf{Descriptive Code and Documentation}: Provides summaries to explain the code’s purpose, functionality, and insights~\cite{pimentel2019large}.
    \item \textbf{Error Handling and Edge Cases}: Handles errors and edge cases gracefully~\cite{perkel2018jupyter}.
\end{enumerate}

The assigned notebooks should evaluated independently by scoring them based on these characteristics. The scoring process is conducted using a numerical scale where a higher score reflects better understandability. The individual scores are then averaged, and a threshold-based approach is employed: If the average score exceeds half of the maximum possible score, the notebook is classified as GCU; otherwise, it is labeled as NCU. The final classification follows a binary scheme (GCU/NCU) to simplify interpretation and provide actionable insights.

For robustness, each notebook is reviewed by three independent annotators, and a consensus approach is followed to ensure reliability. A notebook is classified as GCU only if at least two out of three annotators agree on its classification. In cases of disagreement, additional review steps are taken to refine the assessment.

\paragraph{\textbf{Examples}}
The following are examples of notebooks in each class:

\paragraph{\textit{Examples of NCU:}}
\begin{itemize}
\item https://www.kaggle.com/code/scirpus/standard-gp/notebook
\item https://www.kaggle.com/code/alexandreh13/lung-cancer-simple-analysis-and-classification/notebook
\item https://www.kaggle.com/code/mayer79/m5-forecast-dept-by-dept-and-step-by-step/notebook
\end{itemize}

\paragraph{\textit{Examples of GCU:}}
\begin{itemize}
\item https://www.kaggle.com/code/junkal/breast-cancer-prediction-using-machine-learning/notebook
\item https://www.kaggle.com/code/jneupane12/analysis-of-movielens-dataset-beginner-sanalysis/notebook
\item https://www.kaggle.com/code/sathishrao/iris-data-set/notebook
\end{itemize}

\begin{figure}[t]
    \centering
    \frame{\includegraphics[width=0.7\linewidth]{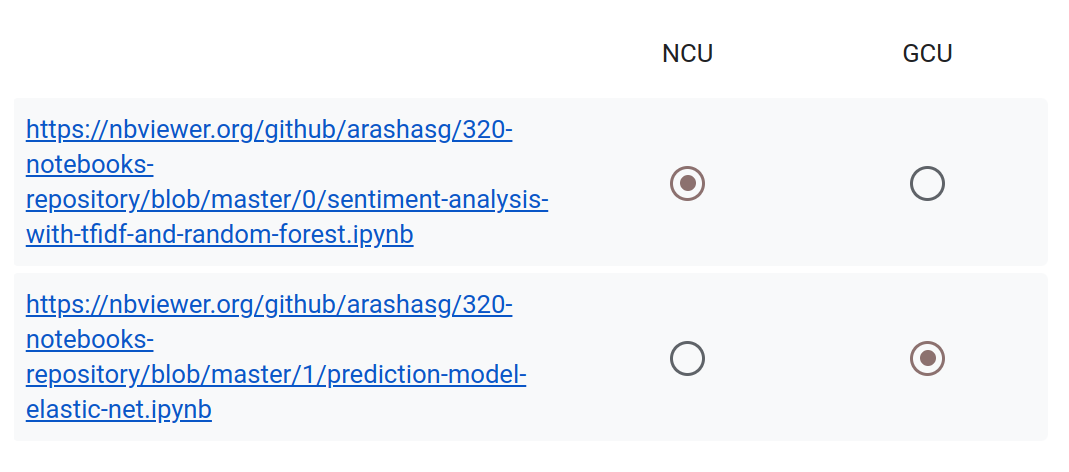}}
    \caption{A sample for classify notebooks}
    \label{fig:annotating_example}
\end{figure}

\paragraph{\textit{Instructions for Annotators}}
To complete the annotation task, please follow these steps:

\begin{enumerate}
\item Access the Google Form link provided to you.
\item Click on each notebook link within the form.
\item Carefully review each notebook, considering the definitions and characteristics mentioned above.
\item Score the notebook for each characteristic based on a numerical scale.
\item If the average score across all characteristics exceeds half of the maximum possible score, label the notebook as ``GCU'' (Good Code Understandability); otherwise, label it as ``NCU'' (Normal Code Understandability.
\item Assign a label to each notebook, as shown in \figref{annotating_example}: either NCU or GCU as per criteria.
\item Submit your responses through the form upon completing the evaluation.
\end{enumerate}

\end{appendices}  

\end{document}